\documentclass[preprint,12pt]{emulateapj}
\usepackage{epsfig}
\usepackage{natbib}
\usepackage{amssymb}
\usepackage{amsbsy}
\usepackage{natbib}
\usepackage[mathcal]{euscript}
\bibpunct{(}{)}{,}{a}{}{,}
\begin{document}
 
\def\today{\number\year\space \ifcase\month\or  January\or February\or
        March\or April\or May\or June\or July\or August\or
September\or
        October\or November\or December\fi\space \number\day}
\def\fraction#1/#2{\leavevmode\kern.1em
 \raise.5ex\hbox{\the\scriptfont0 #1}\kern-.1em
 /\kern-.15em\lower.25ex\hbox{\the\scriptfont0 #2}}
\def\simlt{\vcenter{\hbox{$<$}\offinterlineskip\hbox{$\sim$}}}
\def\simgt{\vcenter{\hbox{$>$}\offinterlineskip\hbox{$\sim$}}}
\def\spose#1{\hbox to 0pt{#1\hss}}
\def\etal{et al.\ }

\title{Young brown dwarfs at high cadence: Warm {\em Spitzer} time series monitoring of very 
low mass $\sigma$~Orionis cluster members}
\author{Ann Marie Cody\altaffilmark{1} and Lynne A. Hillenbrand}
\affil{California Institute of Technology, Department of Astrophysics, MC 249-17, Pasadena, 
CA 91125}
\altaffiltext{1}{amc@astro.caltech.edu}

\begin{abstract}
The continuous temporal coverage and high photometric precision afforded by space observatories has 
opened up new opportunities for the study of variability processes in young stellar cluster members. Of 
particular interest is the phenomenon of deuterium-burning pulsation in brown dwarfs and very-low-mass 
stars, whose existence on 1--4 hours timescales has been proposed but not yet borne out by observations. 
To investigate short-timescale variability in young, low-mass objects, we carried out high-precision, 
high-cadence time series monitoring with the Warm {\em Spitzer} mission on 14 low mass stars and brown 
dwarfs in the $\sim$3~Myr $\sigma$~Orionis cluster. The flux in many of our raw light curves is strongly 
correlated with sub-pixel position and can vary systematically as much as 10\%. We present a new approach 
to disentangle true stellar variability from this ``pixel-phase effect,'' which is more pronounced in 
Warm {\em Spitzer} observations as compared to the cryogenic mission. The light curves after correction 
reveal that most of the sample is devoid of variability down to the few-millimagnitude level, on the 
minute to day timescales probed. However, one exceptional brown dwarf displays erratic brightness changes 
at the 10--15\% level, suggestive of variable obscuration by dusty material. The uninterrupted 24-hour 
datastream and sub-1\% photometric precision enables limits on pulsation in the near-infrared. If this 
phenomenon is present in our light curves, then its amplitude must lie below 2--3 millimagnitudes. In 
addition, we present three field eclipsing binaries and one pulsator for which optical ground-based data 
is also available. 
\end{abstract}


\section{Introduction}

Photometric monitoring of brown dwarfs (BDs) and very low mass stars (VLMSs) in young clusters provides 
insight into the dynamic processes affecting such objects at a few million years of age, including 
accretion, magnetic effects, and star-disk interaction. It has long been known that T~Tauri stars and 
their higher mass counterparts exhibit optical brightness fluctuations of typically a few tenths of a 
magnitude, but in many cases more than a magnitude on timescales of hours to days. However, few 
observations have explored the short-timescale regime (i.e., seconds to hours) in the lowest mass 
objects. This is in part due to the difficulty of obtaining from ground-based facilities suitably long 
time series data with high cadence and minimal interruption. Nevertheless young BDs and VLMSs may display 
significant unexplored variability. Objects with active accretion can exhibit luminosity fluctuations 
related to infalling gas on second to minute timescales, owing to the instability of the shock position 
with respect to the stellar photosphere \citep{2008MNRAS.388..357K,2010A&A...510A..71O}. BDs and VLMSs 
whose surfaces are not obscured by infalling material are expected to display a different type of 
instability-- pulsation fueled by central deuterium burning \citep[e.g.,][hereafter 
PB05]{2005A&A...432L..57P}. The periods for this phenomenon are predicted to be in the 1--4 hour range 
for objects from $\sim$0.02 to 0.1$M_\odot$. Empirical verification of these instability theories through 
detection of short-timescale aperiodic or periodic variability presents a new opportunity to probe the 
properties of young low-mass objects.

A few previous claims of short-period variability in the optical have reported flux changes at the few to 
five percent level \citep[e.g.,][]{2004A&A...419..249S,2001A&A...367..218B,2003A&A...408..663Z}. However, 
our own attempts to identify pulsation from the ground in a number of very-low-mass $\sigma$~Ori objects 
\citep[][hereafter CH10]{2010ApJS..191..389C} resulted in more stringent limits on pulsation and shorter 
timescale periodicities in the optical: If present, $I$-band amplitudes must be less than $\sim$0.01 
magnitudes in the brown dwarf sample, and an order of magnitude lower among the very low mass stars. Thus 
further efforts in the search for and characterization of short-timescale variability require even more 
sensitive observations.

Space telescopes offer a chance for deeper variability searches since the lack of atmosphere minimizes 
systematic errors in photometry, affording signal-to-noise ratios close to the poisson limit. They also 
fulfill the need for dense and continuous time sampling by staring at a single patch of sky for extended 
periods of time without the inconveniences of weather, daytime interruption, or synoptic scheduling. 
Additional progress may be made by observing in the infrared. While this band is not a traditionally 
favored band for photometric time series work, it has several advantages for the detection of 
low-amplitude variability in BDs. Because of their cool temperatures, BDs are brightest at wavelengths 
just longward of 1~$\mu$m and thus should be amenable to relatively high signal-to-noise photometry in 
the near to mid-infrared. Our optical observations revealed that approximately 85\% of low-mass cluster 
members display variability at the 1--10\% level, which we attributed to primarily rotational modulation 
of spots and variable accretion. The amplitude of brightness fluctuations produced by these mechanisms 
is expected to decrease with wavelength \citep[e.g.][]{2009A&A...508.1313F}, thereby reducing confusion 
between pulsation and other sources of variability. Thus while the amplitude range and wavelength 
dependence of pulsation are unknown (the linear stability theory of PB05 predicts only periods, as a 
function of mass), the lower temperature contrast between any magnetic spots or accretion flows and the 
photosphere may enhance the detection probability in the infrared.

We have used the {\em Spitzer} Space Telescope warm mission Infrared Array Camera 
\citep[IRAC;][]{2004ApJS..154...10F,2004ApJS..154....1W} to observe a sample of 14 brown dwarfs and 
low-mass stars in the $\sim$3~Myr $\sigma$~Orionis cluster, several of which had been claimed previously 
as short-period variables \citep[e.g.,][]{2001A&A...367..218B,2003A&A...408..663Z,2004A&A...419..249S}. 
The reported timescales (2--5 hours) are too short to be explained by rotational modulation, suggesting 
that at least some of these objects are good accretion instability or pulsation candidates. With data at 
the 3.6 and 4.5~$\mu$m wavelengths, the pixel-phase effect (i.e., oscillations in the measured flux 
introduced by uncorrected intrapixel sensitivity variations) is more pronounced now than in the 
cryogenic mission. In this paper, we discuss an alternative method relative to the commonly adopted 
approach to remove it. We present light curves sampling timescales from one minute to one day and 
discuss the prospects for pulsation. Finally, we identify several field eclipsing binary and pulsating 
variables for which optical light curves are also available from our ground-based study, and compare the 
behavior at both wavelengths.

\section{Target selection and properties}

Many very-low-mass young cluster members are now catalogued and thus available for time series 
monitoring. Since {\em Spitzer}/IRAC has two fields that are each 5.22$\arcmin$ across, 
only a few clusters in the 1--10~Myr range contain enough known very low mass members to enable 
monitoring of more than one or two BDs simultaneously. Among these are IC~348 and 
$\sigma$~Orionis. We chose the latter for the present study with {\em Spitzer} and the former for 
investigation with the Hubble Space Telescope (results to be presented in a future paper). Within 
$\sigma$~Ori, the objects S~Ori~31, S~Ori~45 have been claimed as short-term variables 
\citep{2001A&A...367..218B,2003A&A...408..663Z}, providing two promising targets for 
our pulsation search. Since the region around them is encompassed by our previously published 
optical fields (CH10), we used the same list of candidate cluster members to 
select targets. We experimented with the position of the 3.6~$\mu$m field as well as its 
orientation with respect to the 4.5~$\mu$m field, whose center is offset by $\sim$6.7$\arcmin$, to 
optimize the pointings.

In addition, we aimed to select targets with a high probability of exhibiting pulsation, based 
on luminosity and temperature consistent with PB05's predicted position of the pulsation 
instability strip. To assess H-R diagram positions, we assembled a set of 50 previously 
confirmed $\sigma$~Ori members with available spectral types, primarily from 
\citet{2003A&A...404..171B}. Temperatures were estimated using the intermediate gravity 
temperature scale derived by \citet{2003ApJ...593.1093L}, which accounts for the lower gravity 
of young objects compared to field dwarfs and is appropriate for the young objects studied here. 
In addition, they have been calibrated for consistency with the \citet{1998A&A...337..403B} 
low-mass evolutionary models, on which the pulsation instability strip from PB05 is based. For a 
few objects without prior spectral types, we obtained low-resolution ($R\sim$1400) spectra from 
the Double Spectrograph on the 200-inch Hale Telescope at Palomar Observatory. Spectral types 
were estimated by comparison with data taken with the same set-up for $\sim$3-Myr low-mass 
IC~348 members previously classified by \citet{2003ApJ...593.1093L}, as well as $\sim$1~Myr 
Taurus and $\sim$5~Myr Upper Scorpius members observed by 
\citet{2006AJ....131.3016S,2006AJ....132.2665S}. Our new spectra, along with those of a number 
of other objects in the $\sigma$~Ori cluster will be presented in a forthcoming paper. We adopt 
uncertainties of 100~K, equivalent to just under one spectral subclass.

Luminosities of $\sigma$ Ori members are dependent upon the estimated distance to the cluster. 
This value has often been taken to be 350$^{+120}_{-90}$~pc, based on the Hipparcos parallax of 
$\sigma$ Ori AB itself. However, \citet{2008AJ....135.1616S} showed that a distance of 
440$^{+30}_{-30}$ is more consistent with main sequence fitting to observations of cluster A 
stars. \citet{2006MNRAS.371L...6J} pointed out that what has traditionally been considered the 
$\sigma$ Ori cluster is in fact likely a superposition of two kinematically distinct groups with 
different radial velocities, ages, and distances. They propose that one of the populations 
corresponds to the Orion OB1a and OB1b association subgroups, while the other is associated with 
the star $\sigma$ Ori itself. With these considerations in mind, we adopt the 
\citet{2008AJ....135.1616S} distance but for completeness we also explore (in $\S$5.1) the effect 
of the smaller value on our computed luminosities and positions on the H-R diagram. The resulting 
distance moduli, $m$-$M$, are 8.21$\pm$0.15 and 7.72 $\pm$0.65 magnitudes. Extinction toward 
$\sigma$ Ori is relatively low, and we adopted A$_{\rm J}$=0.044 \citep{2003A&A...404..171B}.

Final luminosities were determined with $J$-band magnitudes from \citet{2003A&A...404..171B}, 
\citet{2008A&A...478..667C}, and \citet{1999ApJ...521..671B}. Both the $J$ and $I$ bands are
generally favored for their relative lack of contamination from accretion and disk excess. 
However, bolometric corrections in $J$ have the additional advantage of being less sensitive to 
color and surface gravity age \citep[e.g.,][]{1999ApJ...525..466L}. We adopted the bolometric 
corrections used in \citet{2007AJ....134.2340K}, and also used those of 
\citet{2007A&A...470..903C} to check that the results were relatively insensitive to the form of 
the corrections as a function of color and spectral type; we adopt their value of 0.15 
magnitudes as a typical uncertainty.
 
We show the computed locations of our late-type objects on the H-R diagram with respect to the 
theoretical pulsation instability strip in Fig.\ 1, for both possible distance modulus values. 
Uncertainties in luminosity include photometric and bolometric correction errors. However, the 
true errors are dominated by the systematic uncertainty in the distance to the cluster, as shown 
in the figure.

\begin{figure}
\begin{center} 
\epsfxsize=.99\columnwidth
\epsfbox{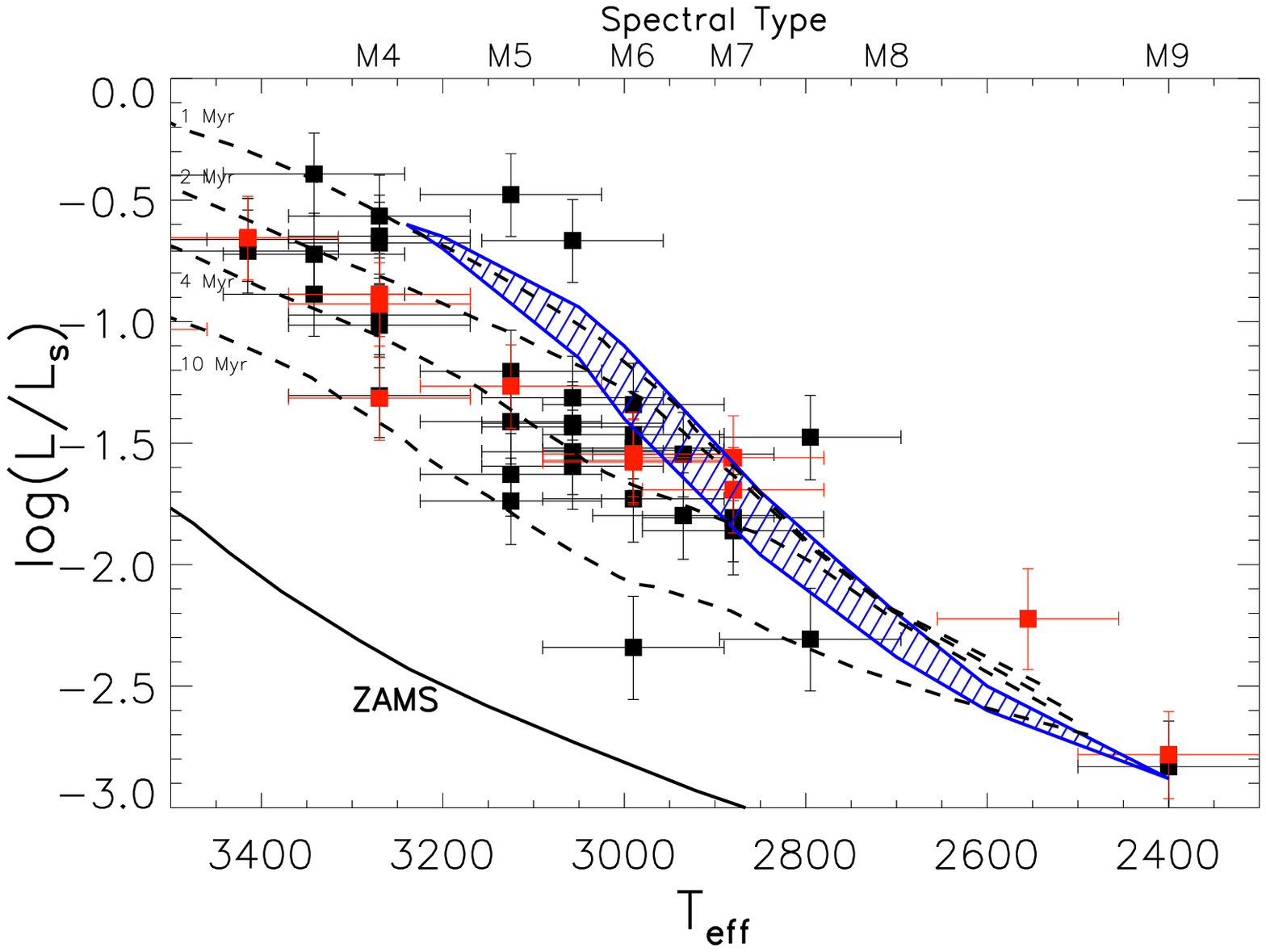}
\epsfxsize=.99\columnwidth
\epsfbox{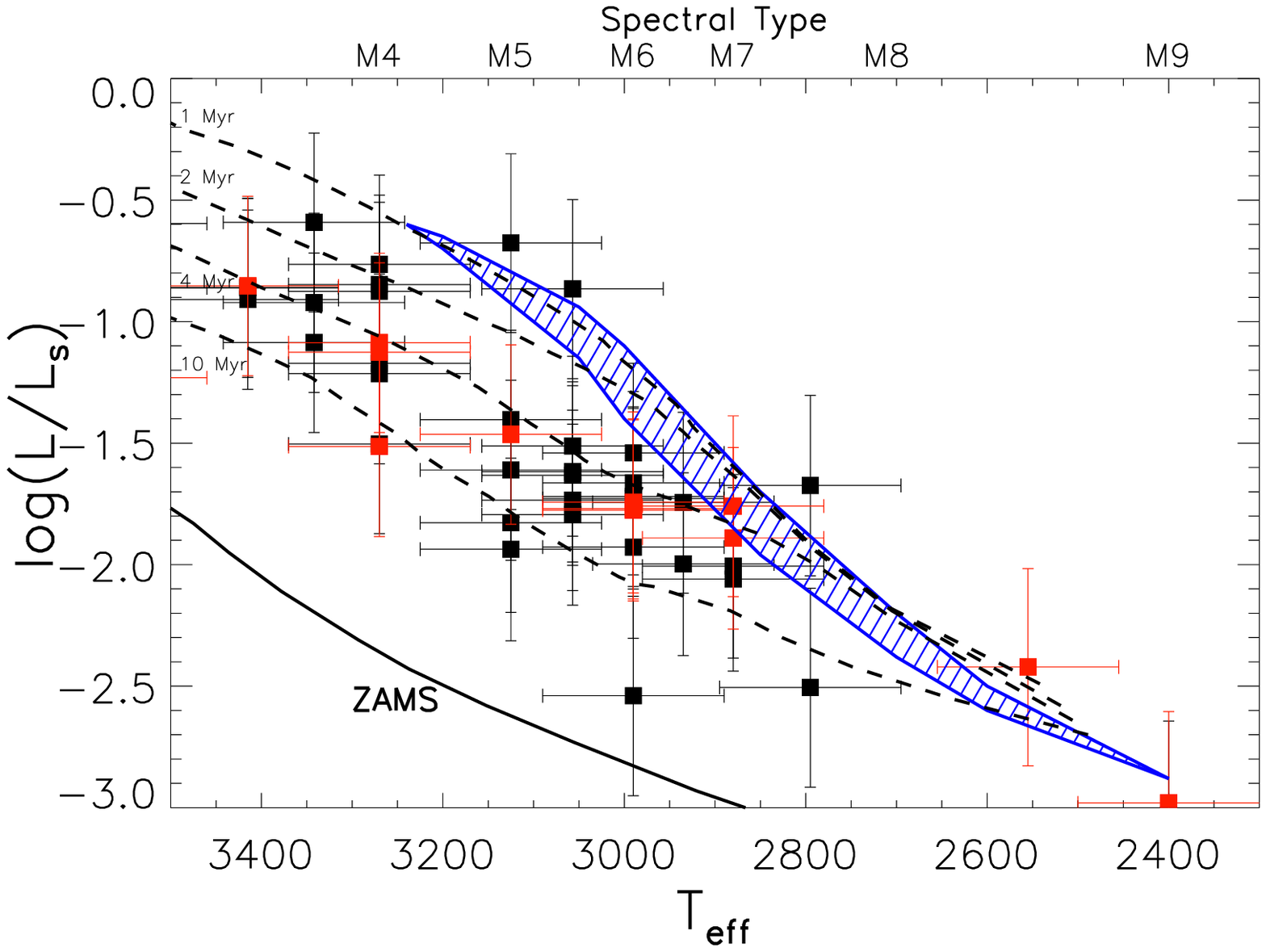}
\end{center}
\caption{The instability strip for pulsating brown
dwarfs and very low mass stars is shown in blue, along with isochrones from
\citet{2003A&A...402..701B} and a sample of spectroscopically confirmed low-mass members of the
$\sigma$~Ori cluster drawn from \citet{2003A&A...404..171B}. In the top panel we assumed a
distance of 440~pc, while in the bottom we changed this to the alternate value of 350~pc.
Targets observed with Spitzer IRAC appear in red.  A number of these objects are on or near the
predicted instability strip, suggesting that they might exhibit pulsation.}
\end{figure}

Additional systematics may be introduced by the choice of band used to calculate the luminosity. 
We performed a comparison test of luminosities derived from the $I$ band for those objects in 
our {\em Spitzer} sample with available $I$-band photometry. There is an 
approximately uniform discrepancy of $\sim$0.35 dex between luminosities derived from the 
$J$-band magnitudes, versus the $I$-band magnitudes. One might conclude that the $J$-band 
magnitudes include contributions from circumstellar disks, but in fact the $J$-band luminosities 
are {\em fainter}. Such a discrepancy may be caused by the unknown difference between the dwarf-like 
bolometric corrections adopted here and those that account for the lower surface gravities of 
young objects. We retain the luminosities as derived from the $J$ band but in $\S$5.1 
compare results from both optical bands.

\footnotetext[1]{http://irsa.ipac.caltech.edu/data/SPITZER/docs/irac/iracinstrumenthandbook/}

The final {\em Spitzer} sample-- shown in red in Fig.\ 1-- included five confirmed and two 
candidate BDs in $\sigma$ Orionis, with three in the 3.6~$\mu$m field and four in the 4.5$\mu$m 
field. In addition, we observed serendipitously seven other known $\sigma$~Ori cluster members in 
the fields which likely are too massive (e.g., $>$0.1~$M_\odot$) to exhibit pulsation but are 
nonetheless valuable targets for investigation of other types of young star variability. This 
brought the total sample to 14 objects-- six in the 3.6~$\mu$m field, and eight in the 
4.5~$\mu$m field. Fewer objects were placed in the 3.6~$\mu$m field because of scheduling 
constraints on the required orientation. Details on each target are provided in Table~1 (see 
Table~1 in CH10 for more details, including coordinates as well as 2MASS identifications). All 
except S~Ori~31 and S~Ori~53 are spectroscopically confirmed members of the $\sigma$~Ori cluster; 
both have colors and spectral type consistent with low-mass $\sigma$~Ori membership, while the 
former also has a proper motion consistent with membership \citep{2009A&A...505.1115L}. In 
addition, we consider object SWW~188 a new spectroscopically confirmed member since it exhibits 
weak Na~I absorption indicative of low surface gravity in the low-resolution spectra that we obtained.

\begin{deluxetable*}{ccccccc}
\tabletypesize{\scriptsize}
\tablecolumns{10}
\tablewidth{0pt}
\tablecaption{\bf Basic target data}
\tablehead{
\colhead{Object}  & \colhead{$I$} & \colhead{3.6} & \colhead{4.5} & \colhead{SpT} & \colhead{Ref} &
\colhead{Optical variability}  \\
}
\startdata
4771-41 & 12.95$\pm$0.02& - & 8.84$\pm$0.02 & K5 & 5 & Aperiodic: RMS=0.23 mags\\
SWW40  & 14.18$\pm$0.03 &- & 11.61$\pm$0.01 & M3 & 5 & Periodic: 4.47d, 0.013 mags\\
S~Ori~J053817.8-024050 & 15.00$\pm$0.04& 11.68$\pm$0.01 & - & M4 & 5 & Periodic: 2.41d, 0.008 mags\\
SWW188 &15.06$\pm$0.03 & - & 12.61$\pm$0.01 & M2 & 5 & - \\
S~Ori~J053823.6-024132 & 15.13$\pm$0.04& 12.17$\pm$0.05 & - & M4 & 5 & Periodic: 1.71d, 0.017 mags\\
S~Ori~J053833.9-024508 & 16.15$\pm$0.04 & - & 12.52$\pm$0.03 & M4 & 5 & Aperiodic: RMS=0.06 mags\\
S~Ori~J053826.8-022846 & 16.17$\pm$0.04 & 12.71$\pm$0.03 & - & M5 & 5 & - \\
S~Ori~J053825.4-024241 & 16.96$\pm$0.04 & 12.96$\pm$0.03 & - & M6 & 2 & Aperiodic: RMS=0.16 mags\\
S~Ori~J053826.1-024041 & 17.05$\pm$0.04 & 13.65$\pm$0.01 & - & M6& 2 & - \\
S~Ori~J053829.0-024847 & 17.06$\pm$0.05 & - & 12.91$\pm$0.03 & M6 & 3 & - \\
S~Ori~27 & 17.22$\pm$0.05 & 13.13$\pm$0.01 & - & M7 & 1 & - \\
S~Ori~31 & 17.46$\pm$0.04 & - & 13.67$\pm$0.02 & M7 & 1 & - \\
S~Ori~45 & 20.03$\pm$0.09 & - & 15.05$\pm$0.05 & M8.5 & 1 & Periodic: 0.3d, 0.034 mags\\
S~Ori~53 & 20.31$\pm$0.09 & - & 17.5$\pm$0.4 & M9 & 4 & -\\
\enddata
\tablecomments{We list the 14 confirmed and candidate $\sigma$~Orionis cluster members
observed with {\em Spitzer}, in order of optical brightness. $I$-band magnitudes are taken from CH10.
3.6 and 4.5~$\mu$m band photometry is the median value determined over our light curves, with  
conservative uncertainties including systematic errors due to poor knowledge of intra-pixel sensitivity
distributions as well as intrinsic variability.  Values listed in the optical variability column are
either the RMS spread of aperiodic light curves over a $\sim$2-week period, or the period and amplitude
of periodic light curves. \\
References-- (1) \citet{2003A&A...404..171B}; (2) \citet{2006A&A...445..143C}; (3) 
\citet{2007A&A...470..903C}; (4) \citet{2001A&A...377L...9B}; (5) 
this work.}
\end{deluxetable*}

\section{Warm {\em Spitzer} observations and data reduction}

Prior predictions and limits on the amplitudes ($\lesssim$0.01 magnitudes) and timescales ($\sim$1--4 
hours) of pulsation guided our observational setup. The ability to detect light curve periodicities at a 
particular amplitude ($A$) depends on both the photometric noise level ($\sigma$) as well as the total 
number of data points ($N$). When identifying a periodic signal in a periodogram (e.g., $\S$5), the 
signal-to-noise ratio in frequency space is roughly equal to $A\sqrt{N}/(2\sigma)$ and must be larger 
than $\sim$4.0 for 99.9\% certainty (see the discussion in $\S$5.1 of CH10). We set a target of several 
millimagnitudes for the minimum detectable periodic amplitude in all objects apart from the faintest two 
BDs.  In addition, data had to be taken frequently enough to probe periodicities on timescales close to 
an hour. Accordingly, observations were carried out over a 24-hour period from 22 to 23 October 2009 
(Astronomical Observation Request key 35146240 and program identification 60169). Exposure times were 
23.6 seconds each, resulting in a cadence of $\sim$30 seconds, and a total of 2730 data points.

Observations at both wavelengths take place simultaneously, with one of the fields in each of the 
3.6~$\mu$m and 4.5~$\mu$m cameras. Therefore, we collected data only in a single band for each of 
our targets. The orientation of the two fields is shown in Fig.\ 2, and the centers were 
R.A.=05$^{\rm h}$38$^{\rm m}$23.3$^{\rm s}$, decl.=-02$\arcdeg$40$\arcmin$29$\arcsec$ (3.6~$\mu$m) 
and R.A.=05$^{\rm h}$38$^{\rm m}$26.4$^{\rm s}$, decl.=-02$\arcdeg$47$\arcmin$13$\arcsec$ 
(4.5~$\mu$m). The position angle was -94.7$\arcdeg$ east of north for both fields. Since our aim was 
to produce photometric time series with as high a precision as possible, we elected not to dither. 
Keeping the positions of all sources fixed within a single pixel reduces the effect of flux 
variations introduced by pixel-to-pixel sensitivity differences not fully corrected by 
flatfielding, although intrapixel sensitivity variation (the ``pixel-phase effect'') remains an 
issue and is addressed below.

For data acquired from the {\em Spitzer}/IRAC camera, all basic calibrations are performed via 
pipeline, as explained in the handbook\footnotemark[1].
As of version 18.12.0, the IRAC pipeline provided images at several different stages of 
processing, from raw unreduced frames to final phot-
\newpage
\noindent ometry ready data. However, at the time of 
writing there were still a number of problems resulting from the transition to the Warm {\em 
Spitzer} mission. Standard bias and dark subtraction, flatfielding, linearity and flux 
calibrations have been applied to create the basic calibrated data (BCD) files. Further 
corrections including automated removal of cosmic rays and the column pull-down effect have been 
performed to create a set of corrected BCDs (CBCDs). Since these procedures were fine-tuned to 
cryogenic mission data, they left numerous column pull-down artifacts as well as a residual bias 
pattern in our data.  Therefore, we elected to carry out the final set of reductions manually, 
starting with the BCDs.

\begin{figure}
\begin{center}
\epsfxsize=0.99\columnwidth
\epsfbox{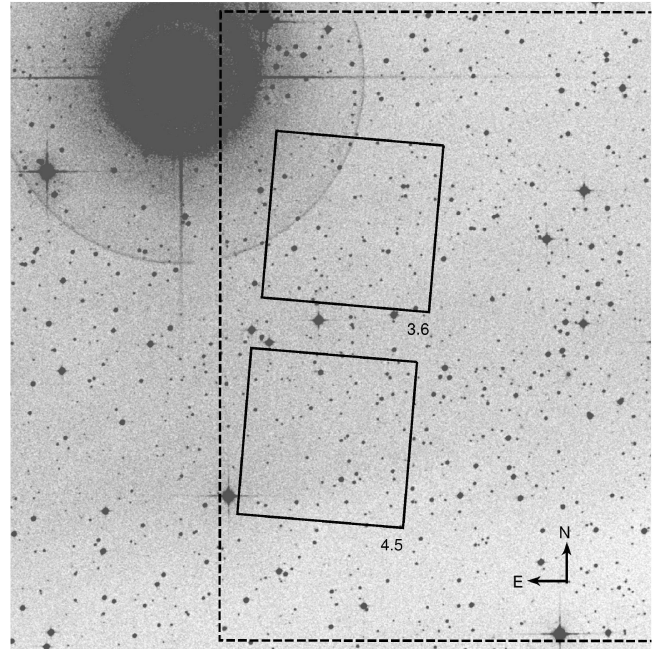}
\end{center}
\caption[]{The {\em Spitzer} IRAC 3.6 and 4.5~$\mu$m fields are overlaid on a Palomar Observatory
Sky Survey 2 (POSS2) red image obtained from the Digitized Sky Survey
(DSS)$^2$. A portion of our 20$\times 20\arcmin$ $I$-band ground-based campaign
with the CTIO 1.0-meter telescope as published by CH10 is also shown for reference (dashed
region).}
\end{figure}


Because there are no laboratory-generated bias frames corresponding to warm mission conditions, 
we retained the bias subtraction applied by the pipeline and modeled the remaining uncorrected 
pattern. Fortunately the residuals largely consist of vertical bands in which brightness 
remained relatively constant throughout our observations. A procedure to median stack all images 
for each channel, mask out the objects, and reset each column to its mean value was performed by 
S. Carey (2010, private communication). Subtraction of the resulting vertical striped bias 
correction image from all BCDs effectively removed the residual patterns.

The column pull-down effect, in which counts are reduced throughout columns with bright 
($>$35,000 DN) sources, was also not fully corrected for in the pipeline. Unlike in cryogenic 
mission data, flux values associated with pull-down now differ above and below the source, in 
addition to following an approximately exponential trend as a function of $y$ position on the 
array. We were provided an updated pulldown correction code (D. Paladini 2010, private 
communication), which satisfactorily modeled and removed this effect.

\section{Photometry routine}

\subsection{Aperture photometry} We performed aperture photometry on our 14 target objects using 
a variety of aperture sizes and sky annulus widths and radii for background subtraction. The IRAC 
camera has a somewhat undersampled point spread function (PSF), with a full-width at half maximum 
(FWHM) size of $\sim$1.4 pixels. As a result, most of the flux from each object is concentrated 
in the central pixel. \footnotetext[2]{http://archive.stsci.edu/cgi-bin/dss\_form} Inaccurate 
aperture centering can thus lead to erroneous brightness 
fluctuations in the resulting light curve. We determined moment centroid positions 
by calculating position-weighted flux averages within a four pixel radius. Points for which the 
centroid algorithm failed due to a cosmic ray or other bad pixel effect were omitted from the 
data (this comprised $\sim$3\% or less of the photometry). Apertures with radii from 
two to four pixels were placed at these centroids, sky annuli of various sizes from 2 to 9 
pixels were used, and the enclosed sky-subtracted flux was determined with the IRAF {\em phot} 
task. We adopted the aperture resulting in the lowest RMS light curve spread, which was 2 pixels 
for most targets. We converted the data to magnitudes by incorporating the published IRAC zero 
point values, aperture corrections, and location-dependent array response provided by the 
handbook. Average 3.6 and 4.5~$\mu$m band magnitudes are listed in Table~1.

Even with careful placement of apertures, many of the light curves contained deviations beyond 
the expected white noise level that were not characteristic of the underlying stellar 
variability.  Points with particularly large flux suggested cosmic ray hits
within the stellar PSF. These occurrences appear random and uncorrelated, and thus unlikely to 
represent real short-term astrophysical behavior. Since we did not dither, it was not possible to 
remove these without binning images or datapoints. We elected instead to filter erroneous flux 
values directly out of the light curves with a 3-sigma clipping algorithm.

\subsection{Pixel phase correction}
Our first pass at the photometry also revealed that most objects suffer from the well known 
IRAC pixel-phase effect: although target positions were restricted to a single pixel, 
movement of the centroid within the pixel introduces position-correlated flux changes of up 
to 10\% due to response variations within individual pixels 
\citep{2006ApJ...653.1454M,2011ApJ...726...95D}. The $x$ and $y$ centroid positions 
executed not only several small jumps, but also an oscillatory motion with period $\sim$60 
minutes due to the subtle effect of a thermal cycling battery heater on {\em Spitzer} 
pointing. As a result, most of the light curves from channel 1 exhibited periodic fluctuations of 
up to $\sim$4\% amplitude, along with additional systematics of up to 10\%, or 0.1 
magnitudes. We display a typical example of $x$ and $y$ trends as a function of time in Fig.\ 3. The 
effect is about half as large in channel 2 but still significant enough to require removal in many of the 
light curves.

\begin{figure}
\begin{center}
\epsfxsize=0.99\columnwidth 
\epsfbox{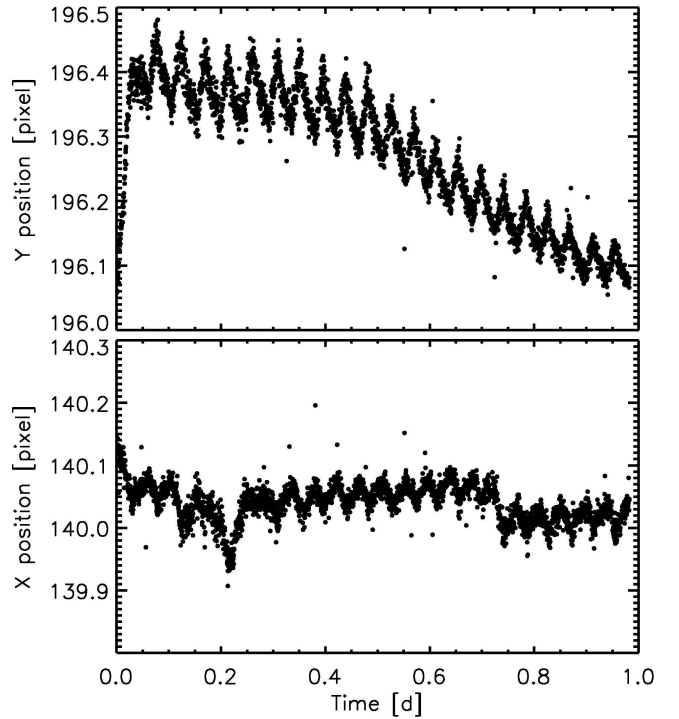}
\end{center}
\caption{$X$ and $Y$ pixel centroid positions of one of our targets (S~Ori~J053817.8-024050)
as a function of time. Since telescope pointing affects all targets in a similar fashion, both the
short-timescale ($\sim$0.04d) oscillations and the more systematic trends are typical of the
centroid behavior of other observed objects as well. Outlier points indicate where the
centroiding algorithm has failed (e.g., because of a cosmic ray hit or other artifact).}
\end{figure}

The Warm {\em Spitzer} mission guide presents a method to correct these effects by 
providing a model of the sensitivity variations within a pixel. However, the model was 
derived from observations of a single bright star and does 
not account well for differences in the response patterns of different pixels. We found that the 
proposed correction algorithm was not adequate for removing the pixel-phase related noise from our 
light curves. Typical signal-to-noise ratios were 55--60\% of that estimated based on the 
Poisson limit, whereas previous work with warm {\em Spitzer} data suggests that we should 
be able to achieve upwards of 75--80\% \citep{2011ApJ...726...95D}. On the other hand, 
subtraction of a median-fit trend from each light curve confirmed that the white noise 
level did indeed reach a level consistent with these predictions.

To recover the additional $\sim$20\% in S/N, we explored several methods for removing noise 
due to the pixel-phase effect. The failure of the model based on a single bright star implied 
that the spatial response differs significantly from pixel to pixel. Therefore, we attempted 
to fit each object's flux with polynomials as a function of $x$ and $y$ position. 
Unfortunately this approach proved problematic for several reasons. First, the pointing during 
our Astronomical Observation Request (AOR) traced out a region in $x$-$y$ space that was 
neither homogeneous nor large compared to the pixel size (e.g., Fig.\ 4). Rather, small 
pointing jumps led to centroid positions occupying three somewhat discrete areas of the pixel. 
In addition, we were concerned that intrinsic variability of our young cluster sources could 
complicate the fitting process.

\begin{figure}
\begin{center}
\epsfxsize=0.99\columnwidth 
\epsfbox{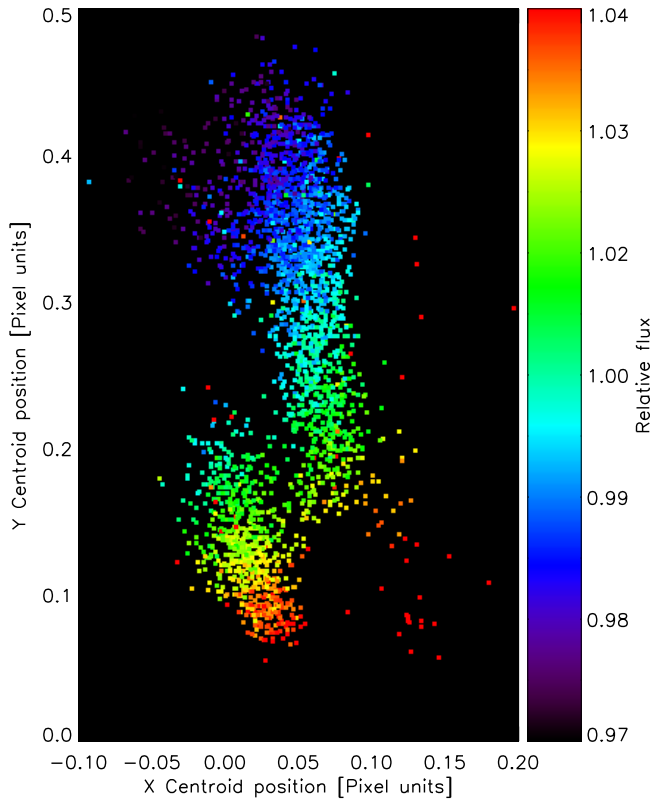}
\end{center}
\caption{Variations in the flux of object S~Ori~J053817.8-024050 measured within a single pixel. The
pixel is centered at [0,0] and extends to $\pm$0.5 in the $x$ and $y$ directions. Only a portion
of the pixel is depicted here.}
\end{figure}

Plots of flux versus $x$, $y$, and phase (distance from a fixed point near the center of the pixel) did 
not exhibit tight trends, suggesting that accurate removal of systematic effects would not be feasible. 
As a result, we opted to fit a gaussian functional form to each object's spatial flux distribution. The 
Warm {\em Spitzer} guide\footnote[3]{http://ssc.spitzer.edu/irac/warmfeatures} suggests that a double 
gaussian function (i.e., sum of gaussians in the $x$ and $y$ directions) is the best-fitting pixel 
sensitivity model based on bright star data. However, because of the incomplete spatial coverage within 
each pixel, we suspected that a single gaussian with adjustable center would work as well. Our adopted 
pixel sensitivity model thus consisted of four free parameters: $$\Delta F 
e^{-((x-x_0)^2+(y-y_0)^2)/2\sigma^2}+F_0,$$ where $\Delta F$ is the height of the gaussian function, $x$ 
and $y$ are the positions with respect to the center of the pixel, $x_0$ and $y_0$ are the offsets of the 
peak flux response from the center of the pixel, and $\sigma$ is the width of the gaussian. $F_0$ is a 
constant determined so that the function averages to 1.0 over the entire pixel.

To identify the best-fitting pixel-phase function we created a script to iterate through a 
reasonable range in the four parameters, perform the pixel phase correction based on each 
set, and assess the presence of pixel-phase noise in the resulting light curve. This 
assessment was performed by generating a periodogram in the range of frequency space where 
the pixel-phase oscillation dominates: 21.5--25 cycles/day (corresponding to periods of 
$\sim$1--1.1 hours, and unfortunately very close to the pulsation signature that we seek). 
It is here that a large peak is seen in the periodograms of raw light curves (Fig.\ 5). We 
present as the ``corrected'' light curve the one for which the integrated periodogram 
in this region is minimized. In some cases (S~Ori~31, S~Ori~J053833.9-024508, SWW~188, 
S~Ori~45, and S~Ori~53), the initial periodogram did not display a peak associated with the 
pixel-phase effect, and so we did not apply any correction. Since the correction process only 
targets a small region of frequency space in the periodogram, it should preserve variability that 
is intrinsic to the objects, if present.

We emphasize that we have chosen the symmetric gaussian pixel-phase model out of 
convenience and lack of knowledge of the underlying distribution; the true pixel 
sensitivity function is likely to be much more complicated 
\citep[e.g.,][]{2010PASP..122.1341B}. The presented light curves may thus have systematic 
inaccuracies. In addition, since the correction process removes only variation on the known 
$\sim$1-hour period of the thermal oscillation, it is not obvious as to whether variation 
on longer timescales is intrinsic to the sources or undercorrection of the pixel-phase 
effect. We caution that any Warm {\em Spitzer} studies attempting to assess variability 
whose precise nature (i.e., light curve shape) is not known in advance will face this issue.

\section{Variability search results}

Our main aim in analyzing the light curves of low-mass $\sigma$~Ori cluster members is to 
search for periodicities on the 1--4-hour timescales predicted for deuterium-burning 
instability. We produced fourier transform periodograms \citep{1975Ap&SS..36..137D} for 
both the raw and pixel-phase-corrected light curves; all are presented in Fig.\ 5. Since 
the observations were continuous over a 24-hour period at 30-second cadence, we are 
sensitive to periodic variability on timescales from one minute to one day. In addition, 
the periodogram does not suffer from aliasing, so true signals are relatively easy to 
identify if they rise high enough above the noise baseline. In most cases, the periodograms 
display a relatively uniform mean from frequencies at a few cycles per day (cd$^{-1}$) out 
to the Nyquist limit at 1440 cd$^{-1}$.  This white noise level depends on the magnitude of 
the source and ranged from 0.001 to 0.004 magnitudes in the periodogram.

\begin{figure*}
\begin{center}
\epsscale{1.15}
\plottwo{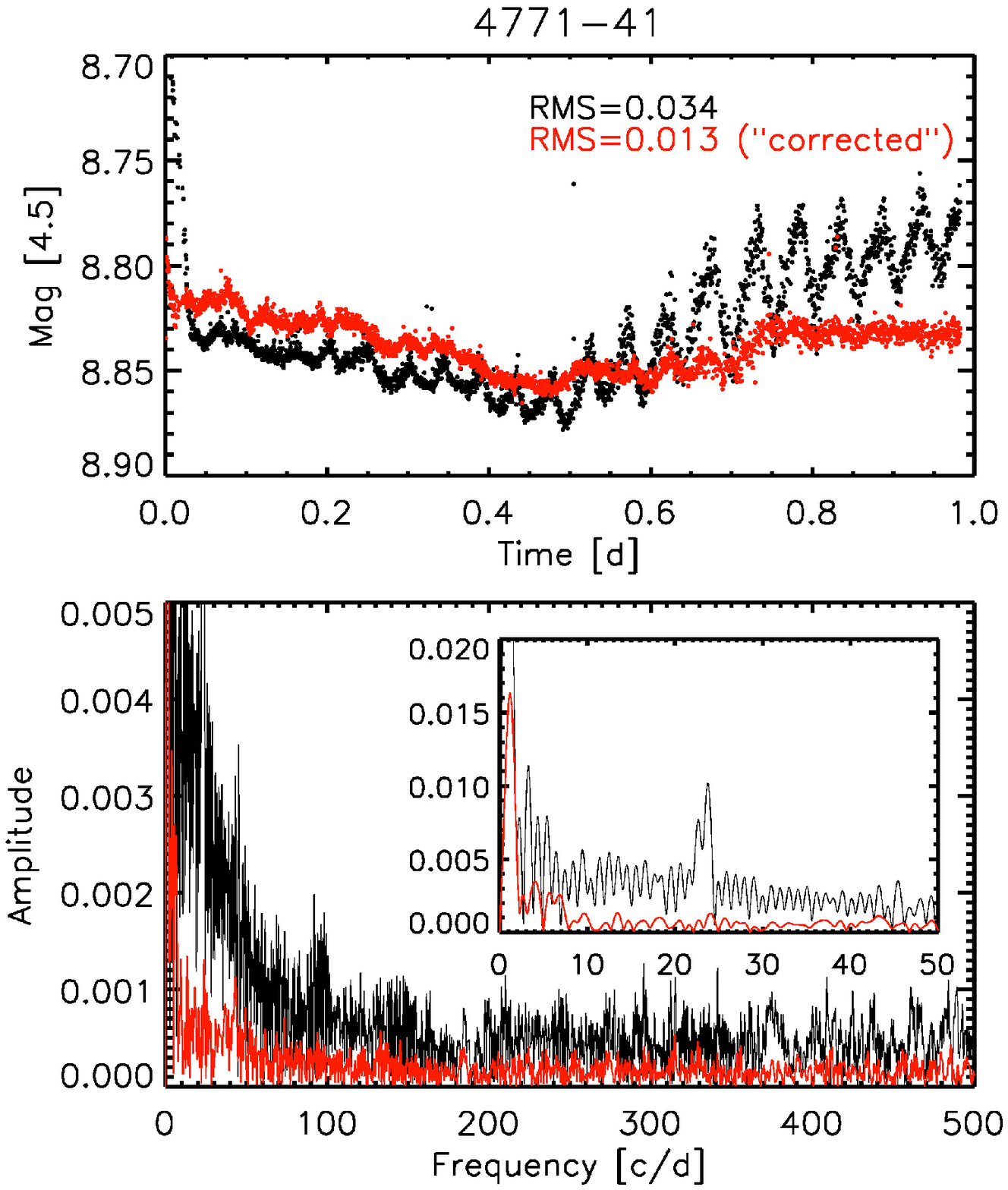}{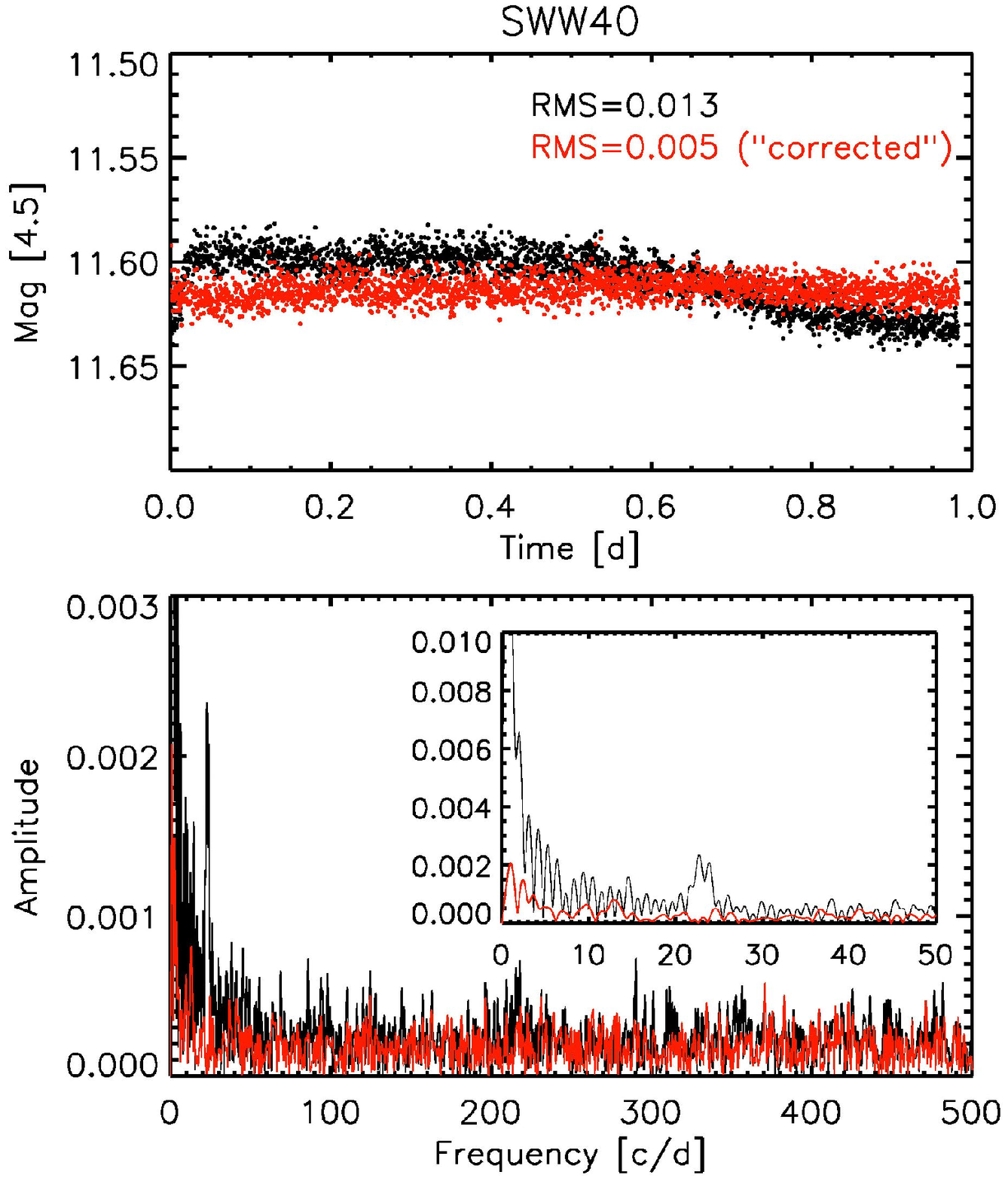}
\vspace{-1.cm}
\plottwo{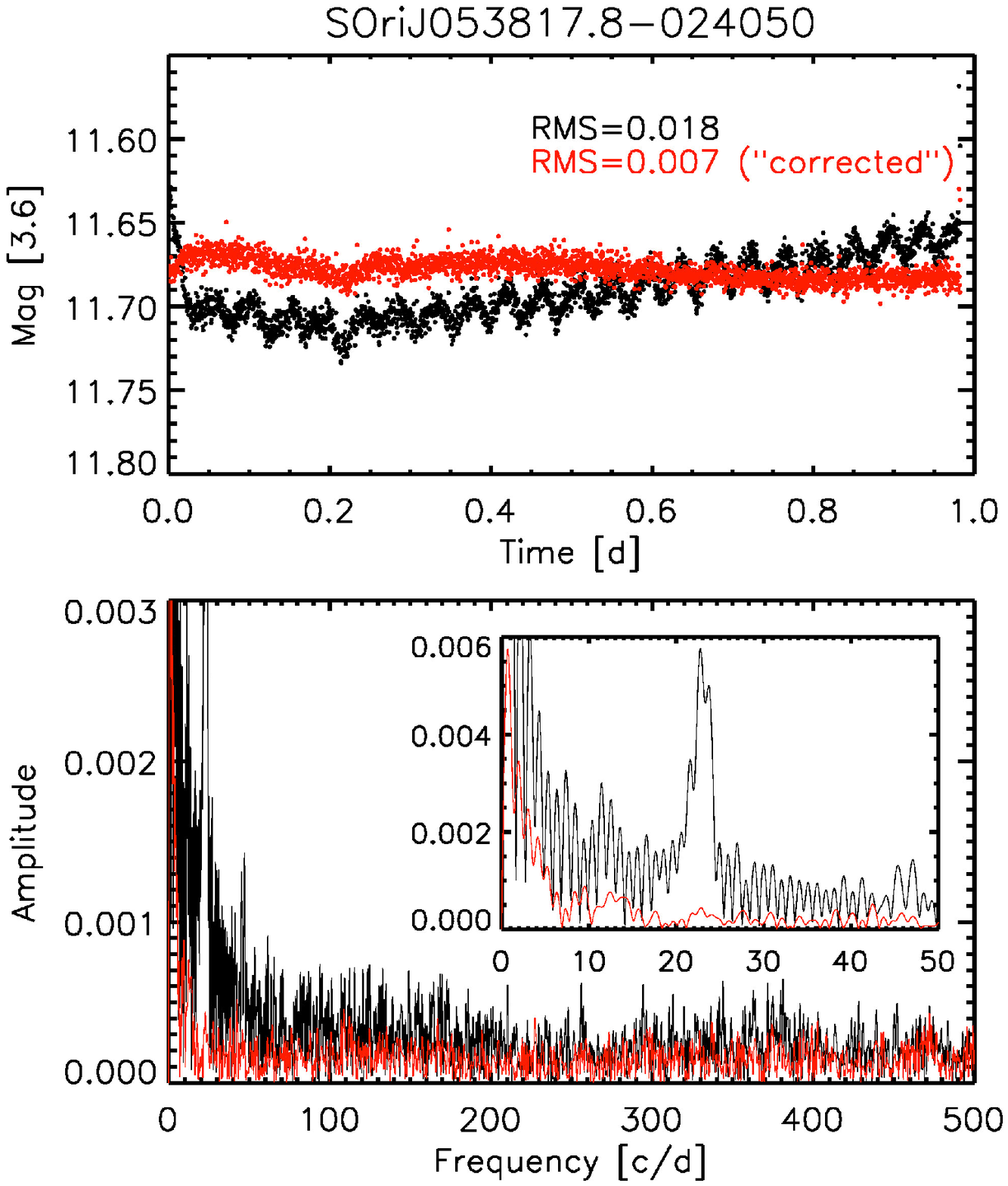}{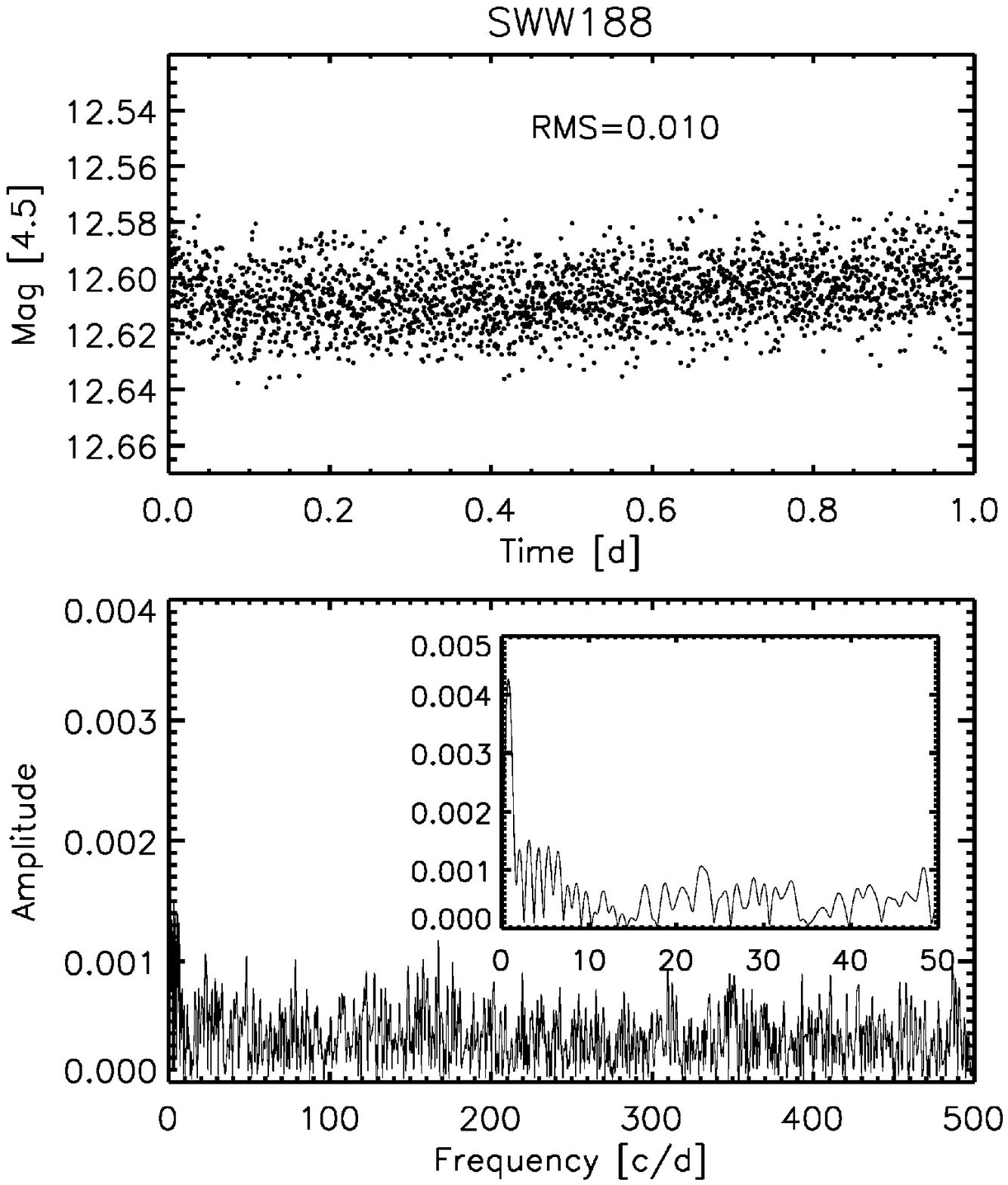}
\end{center}
\caption{Light curves and periodograms for all targets, in order of decreasing optical brightness.
Object identifications are listed above each light curve, and the {\em Spitzer} band (3.6 or
4.5~$\mu$m) is noted in the y-axis label. Black indicates the raw light curve and periodogram,
whereas red shows them after correction for intra-pixel sensitivity effects. Objects with no red
points did not require correction. Periodogram frequencies are given in cycles per day. Insets  
show the same periodograms zoomed in to the low-frequency range where the
signature of the pixel-phase oscillation is visible ($\sim$22-24 cd$^{-1}$).}
\end{figure*}

\addtocounter{figure}{-1}
\begin{figure*}
\begin{center}
\epsscale{1.15}
\plottwo{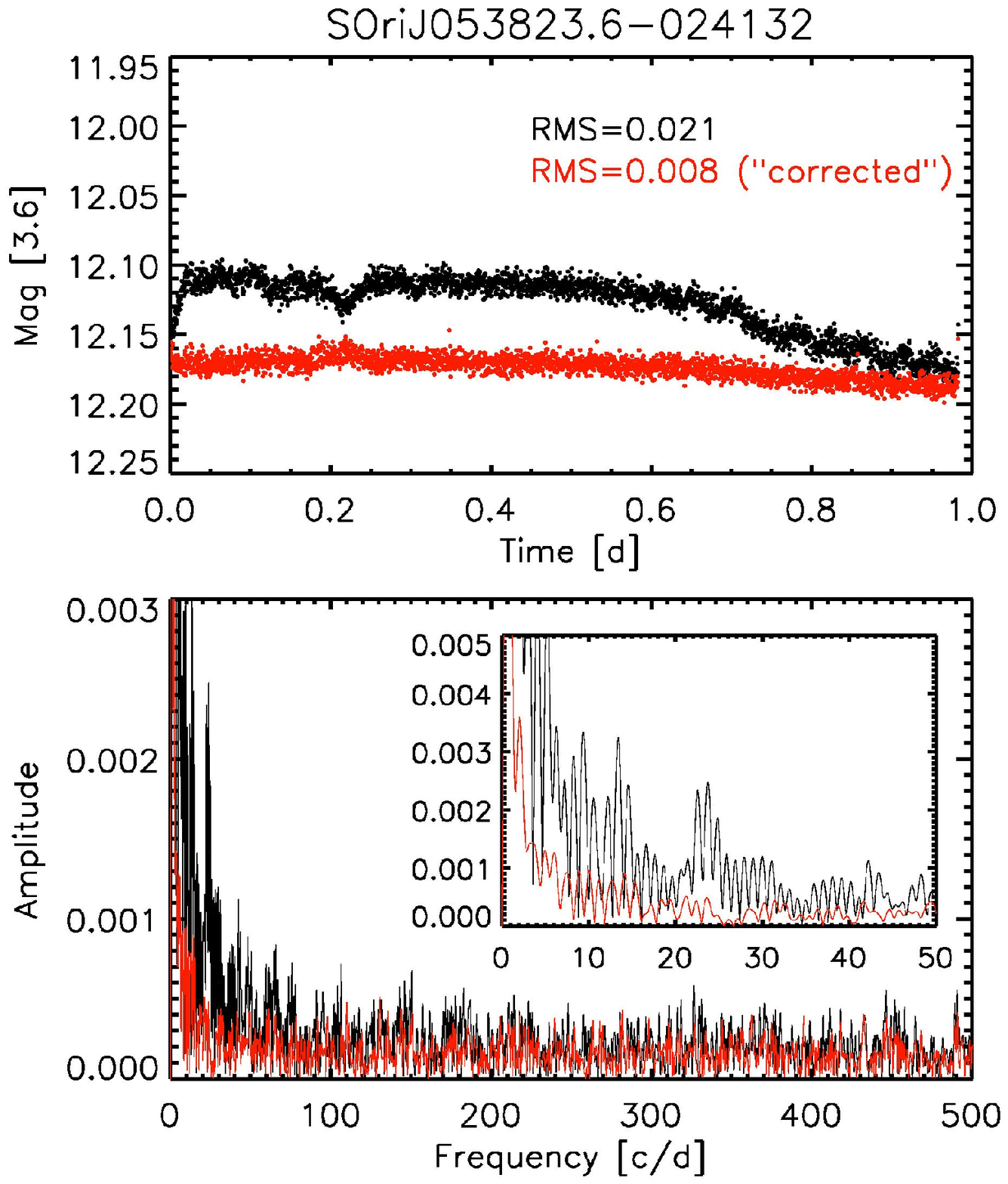}{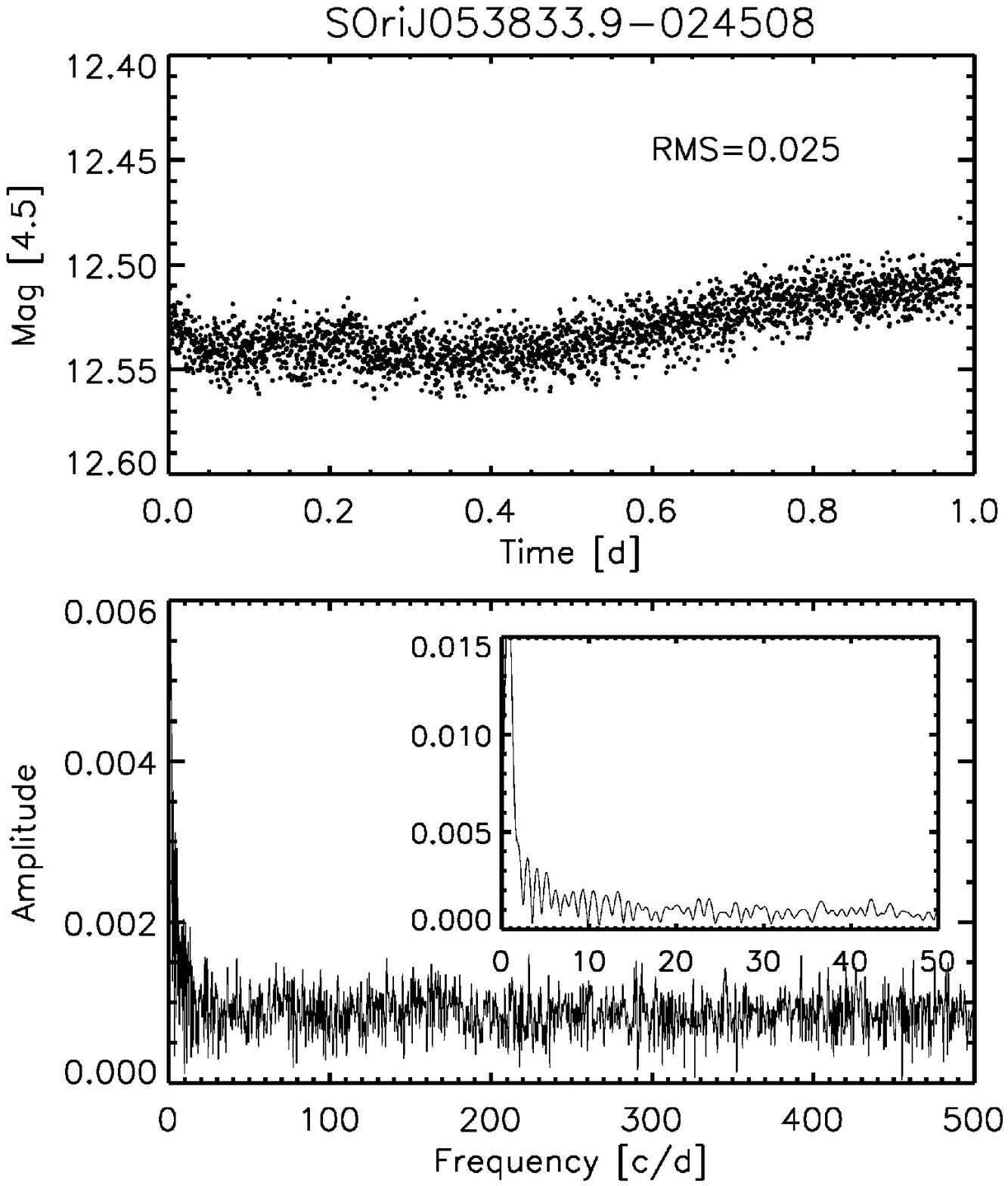}
\vspace{-1.cm}
\plottwo{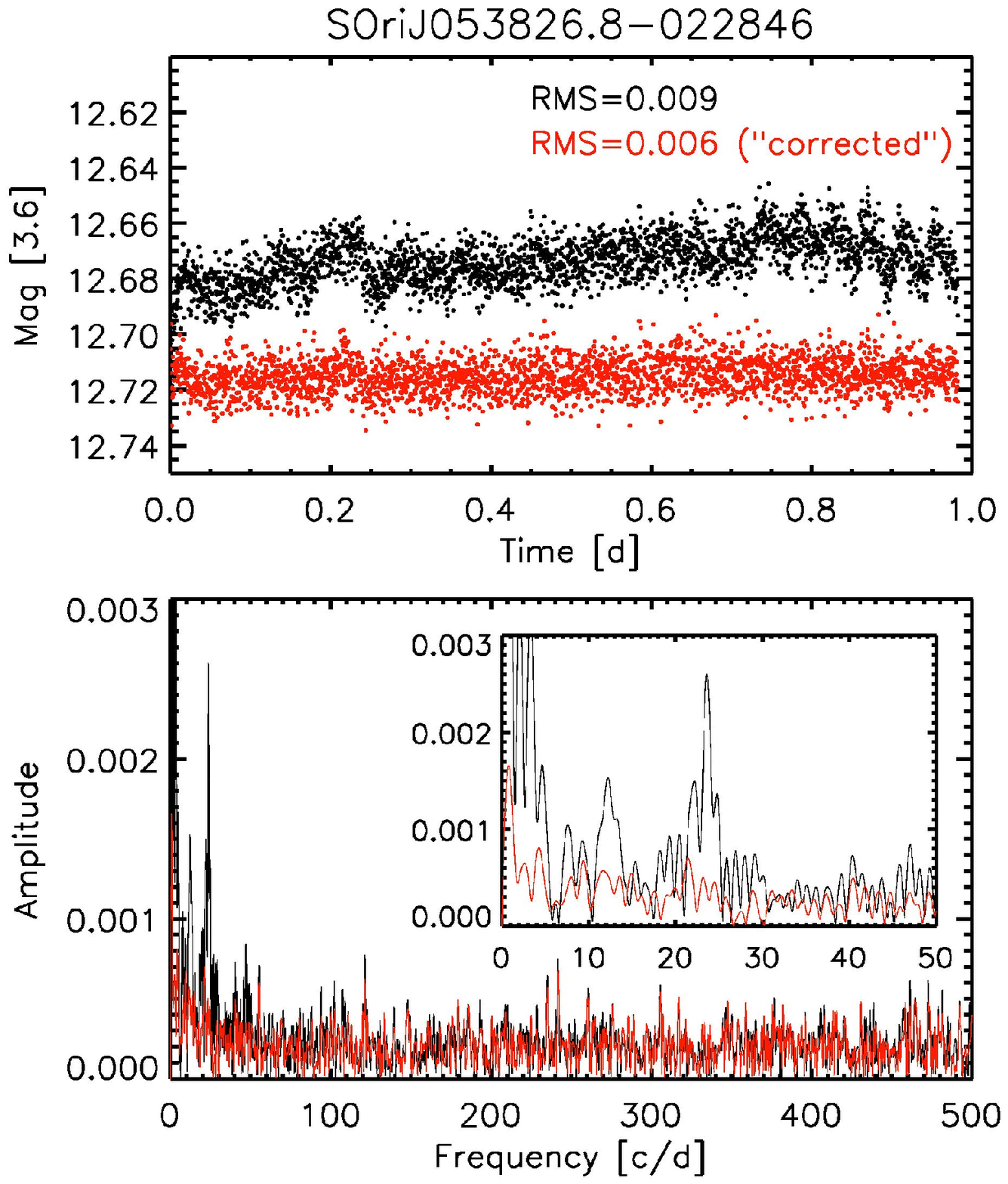}{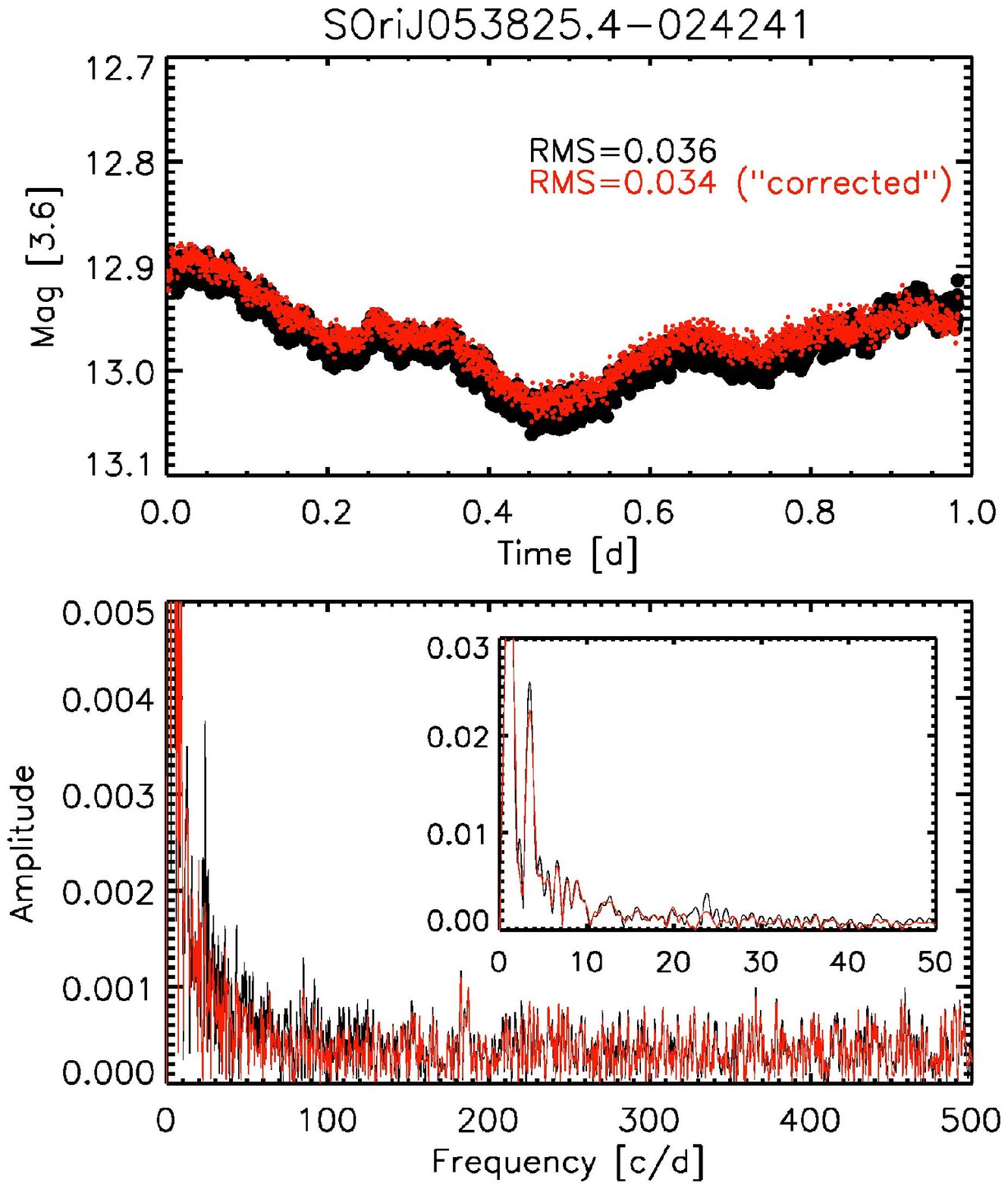}
\end{center}
\caption{(cont.)}
\end{figure*}

\addtocounter{figure}{-1}
\begin{figure*}
\begin{center}
\epsscale{1.15}
\plottwo{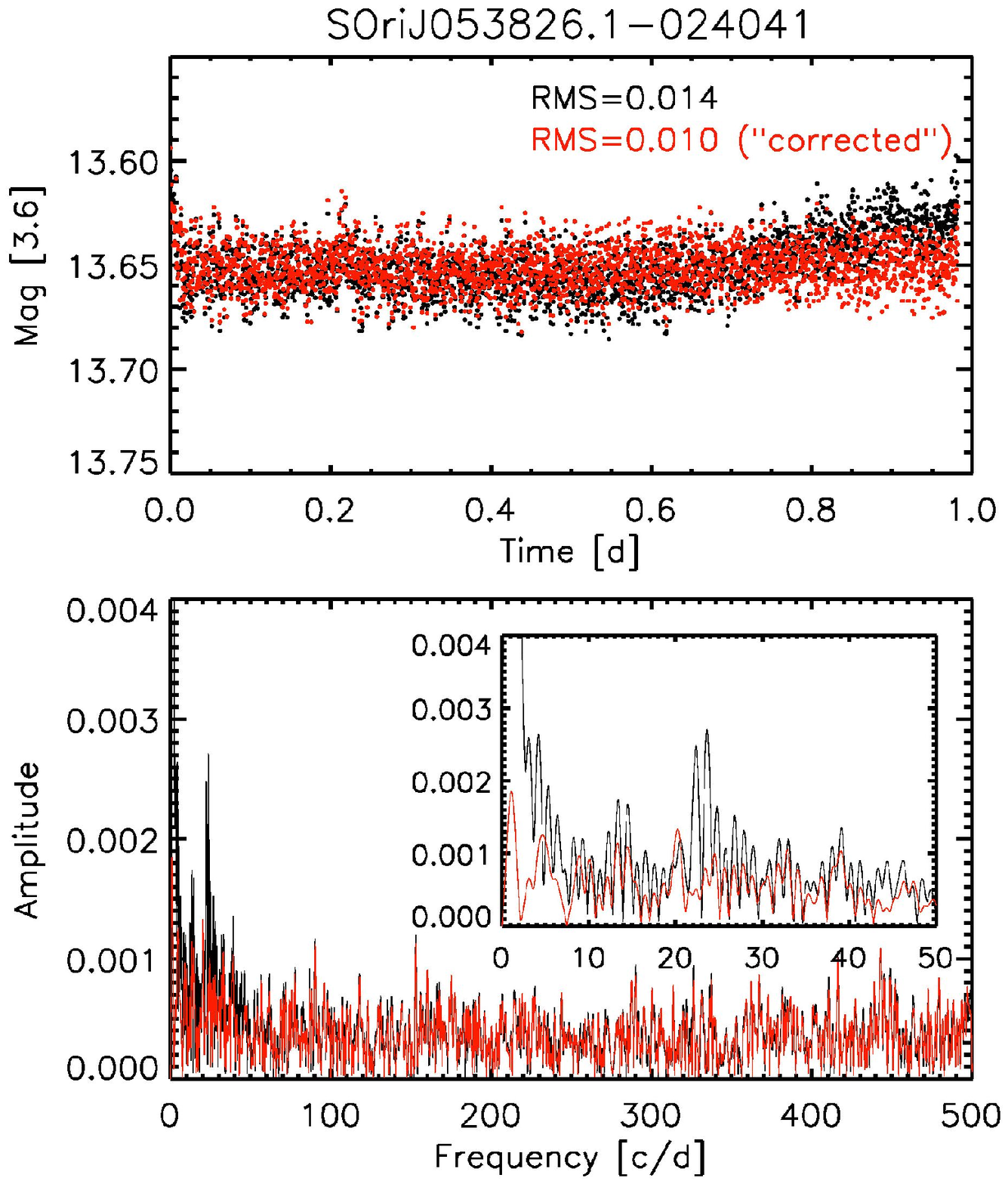}{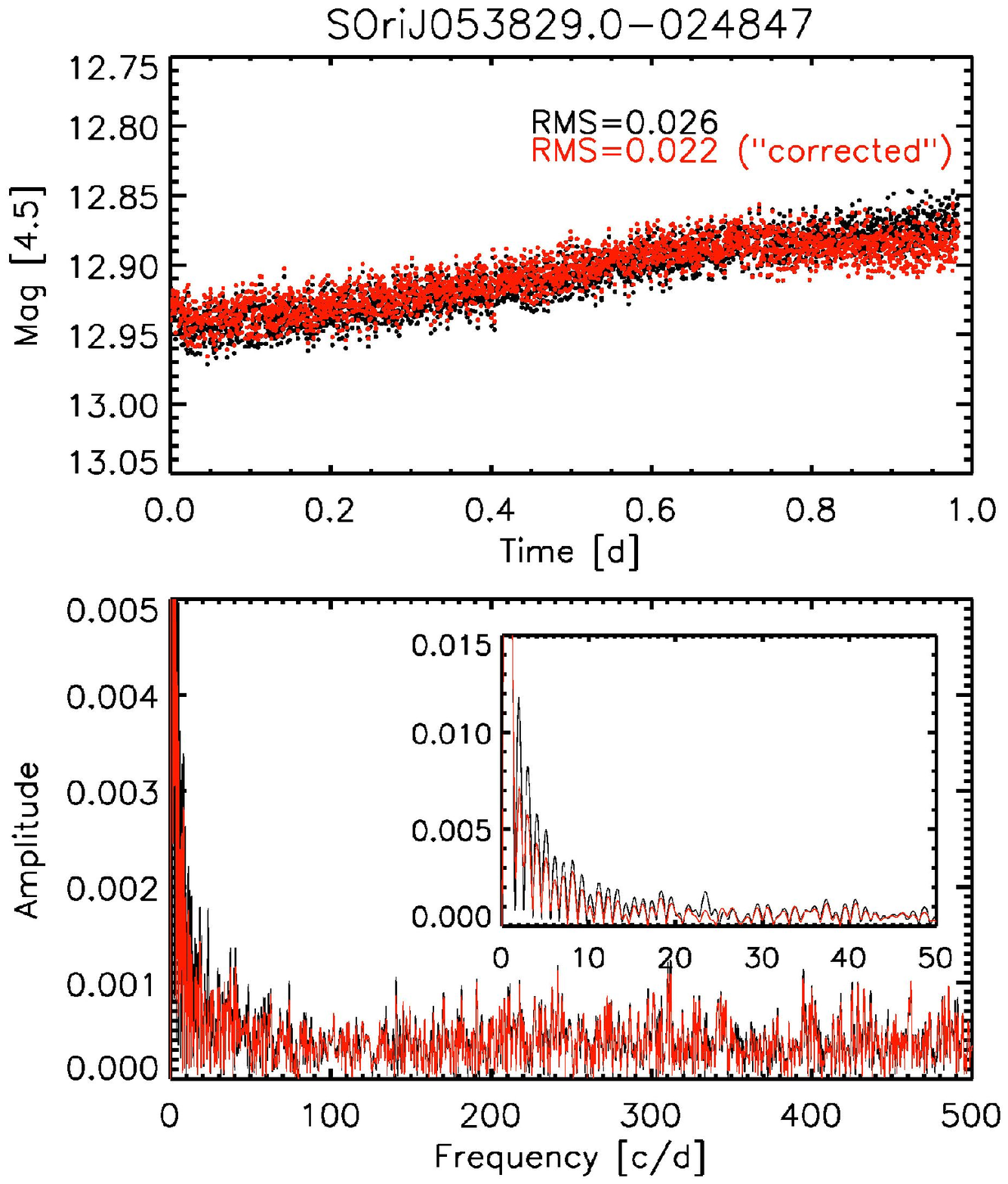}
\vspace{-1.cm}
\plottwo{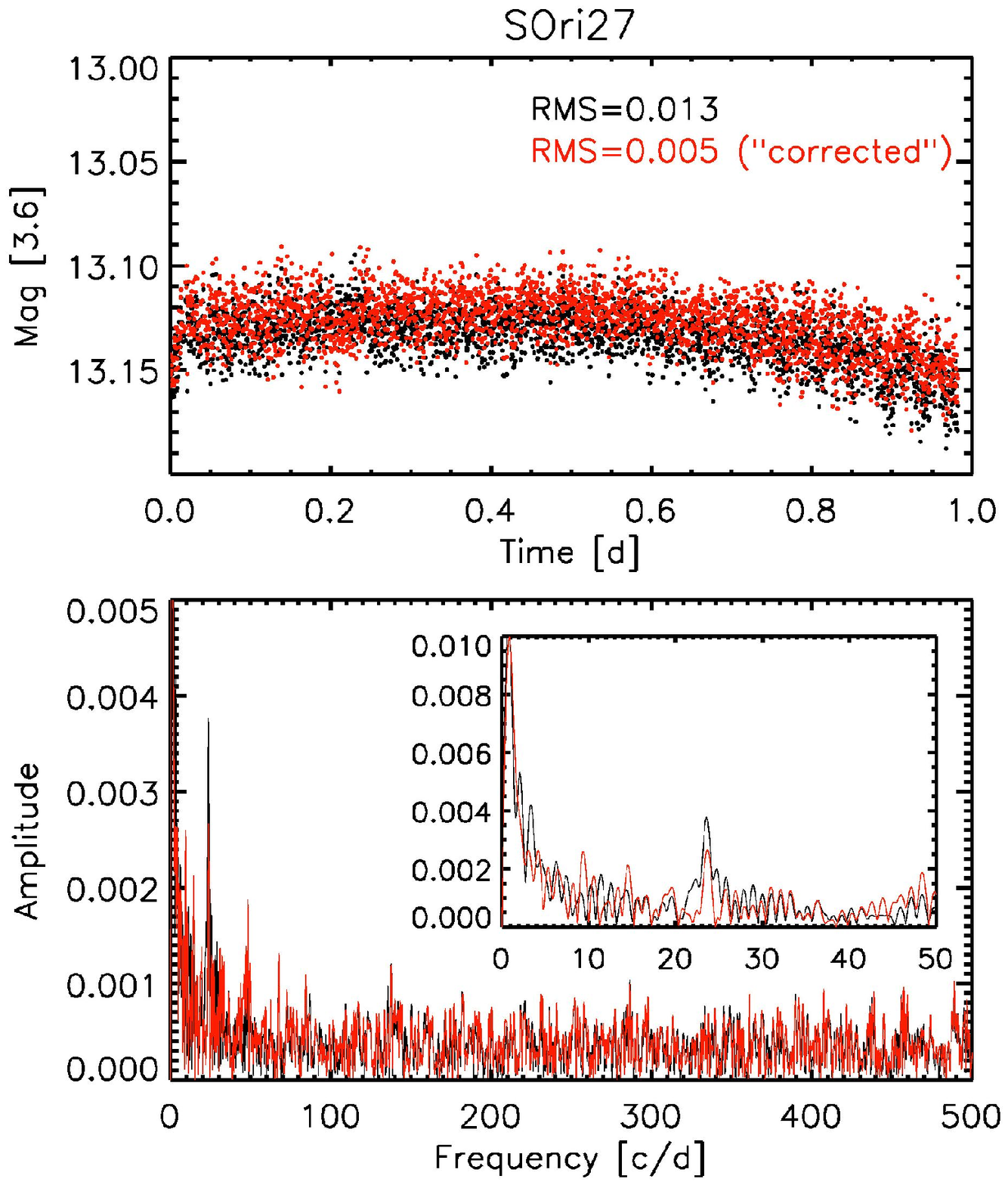}{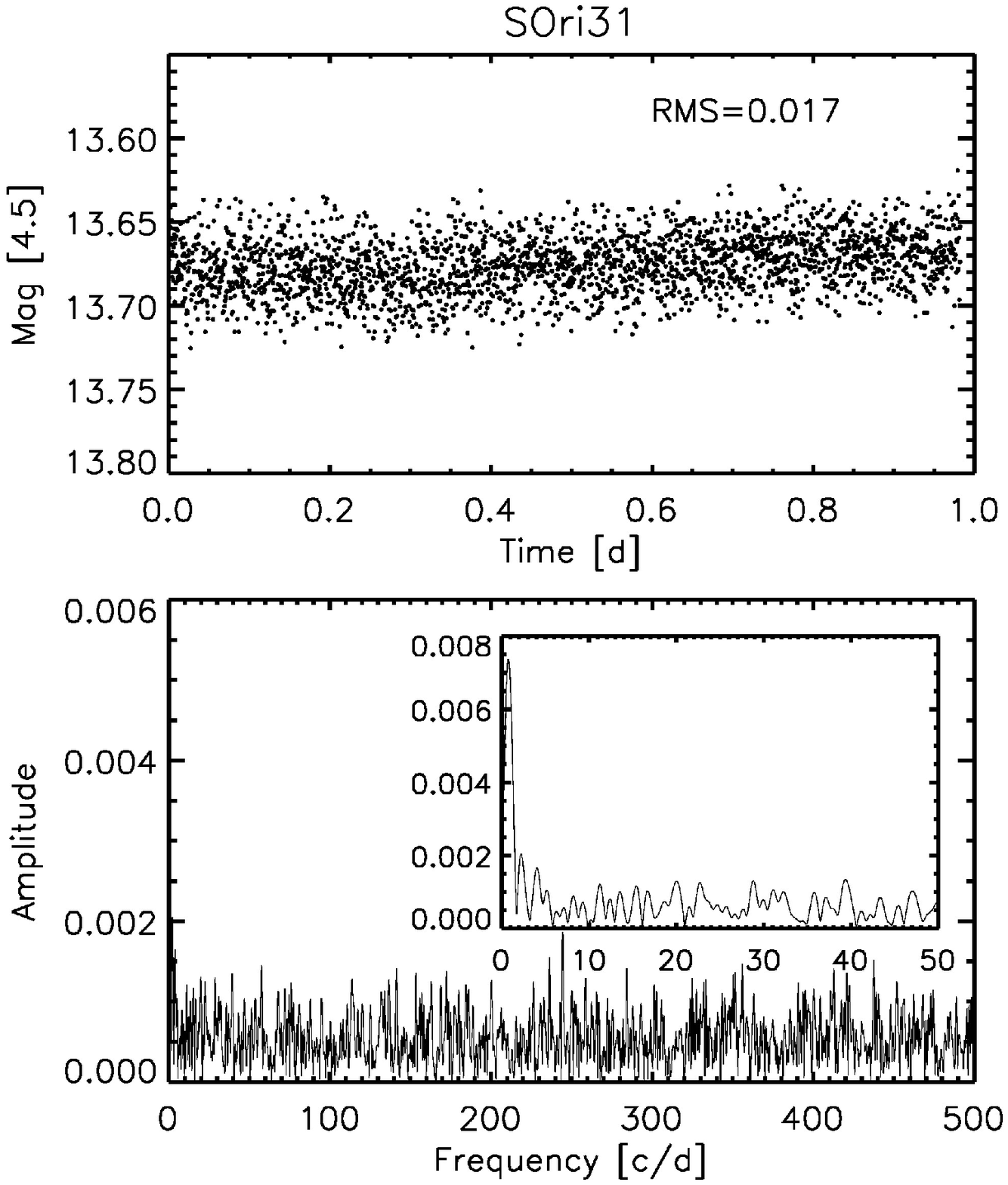}
\end{center}
\caption{(cont.)}
\end{figure*}

\addtocounter{figure}{-1}
\begin{figure*}
\begin{center}
\epsscale{1.15}
\plottwo{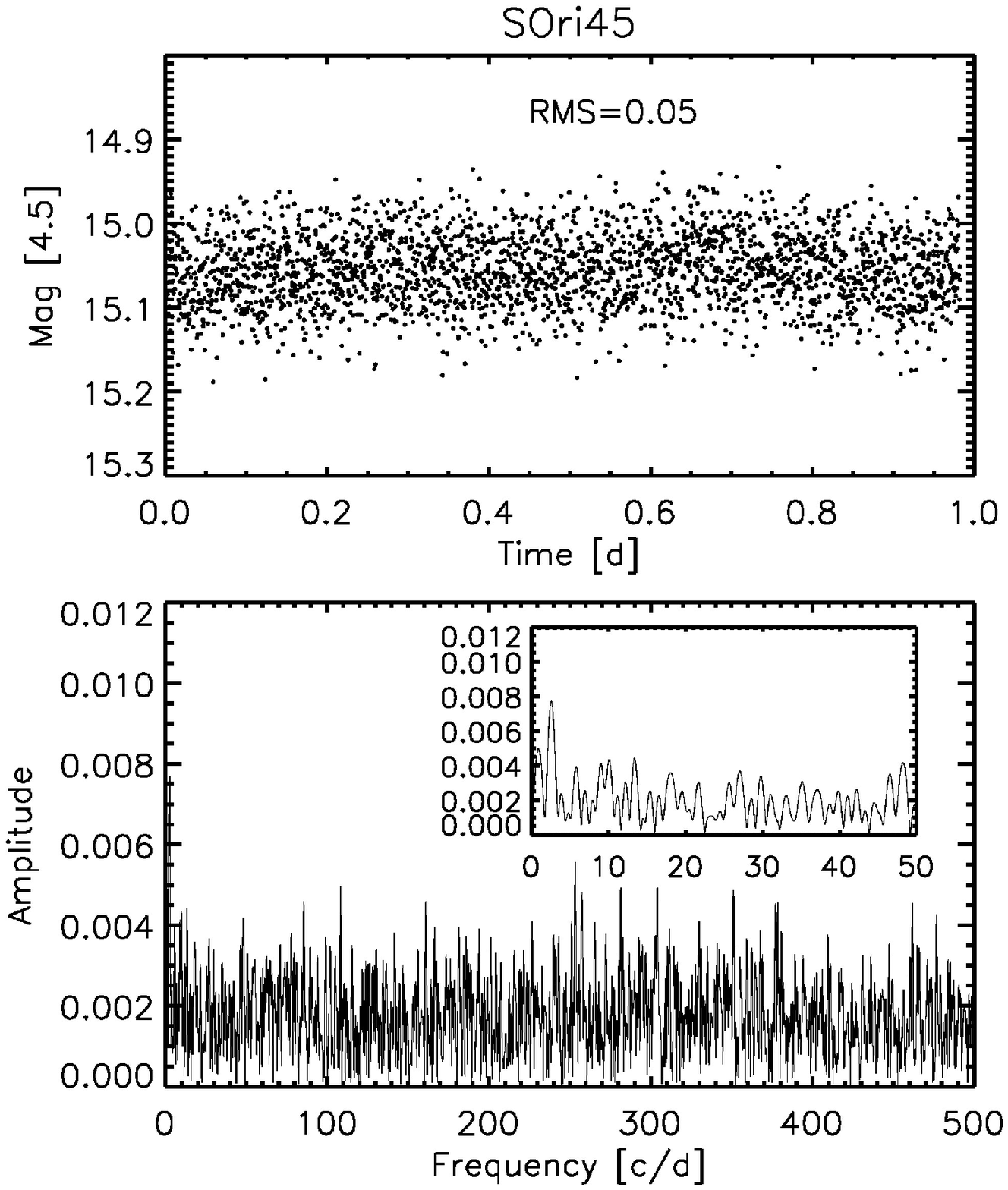}{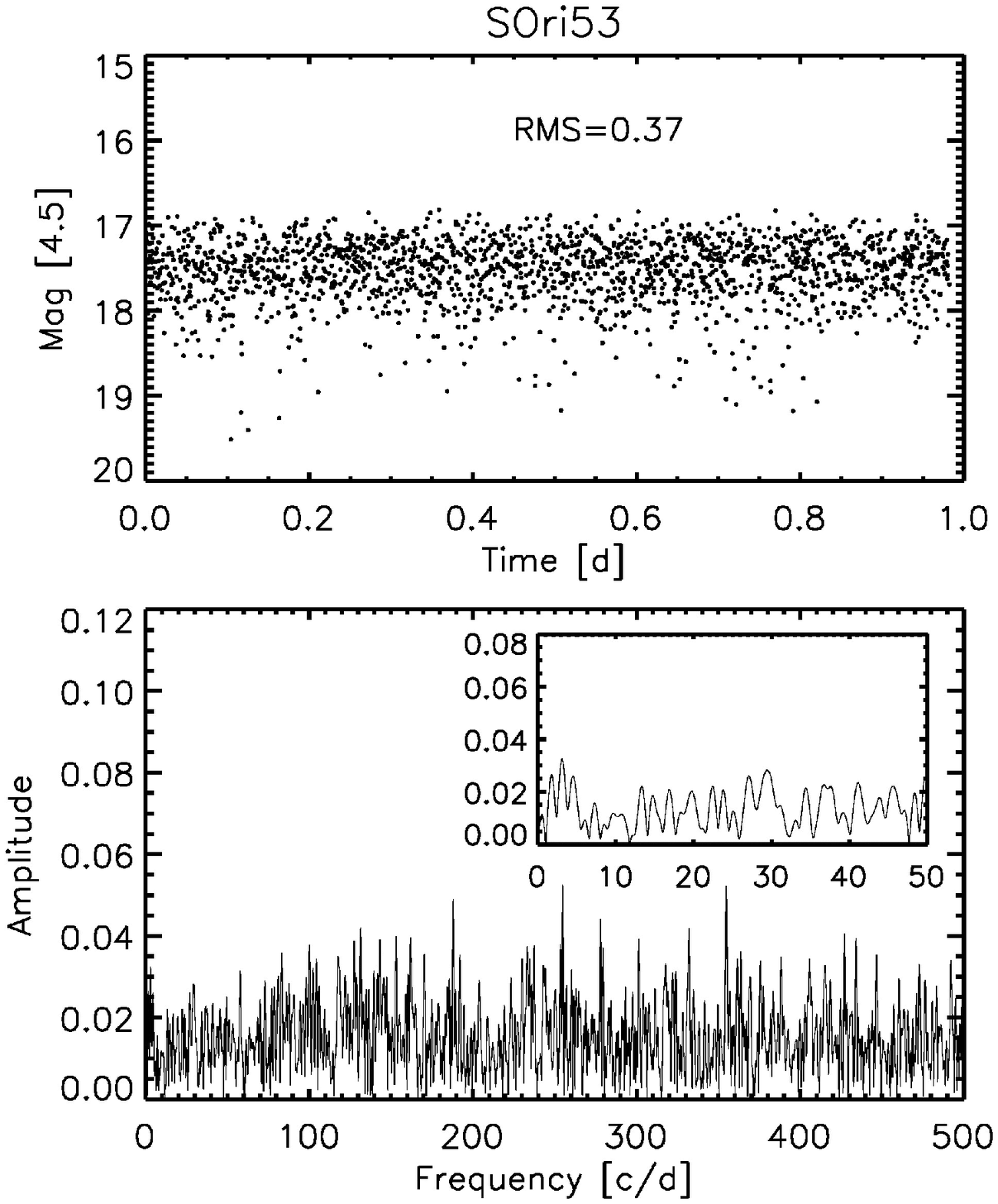}
\end{center}
\caption{(cont.)}
\end{figure*}

Examination of the periodograms revealed that the pixel-phase correction process 
substantially lowered the noise level, enabling better sensitivity to periodicities outside 
the 1--1.1-hour range of the pixel-phase oscillation. The two exceptions were SOri~27 and 
4771-41. The former was centered near the edge of two pixels, making a fit to the spatial 
distributions difficult without resorting to a more complex non-gaussian function. Object 
4771-41 is exceedingly bright, and residual variability seen in the final light curve may be 
a figment of the correction process.

The majority of periodograms are relatively featureless, reflecting minimal variability in 
the input light curves. In the low-frequency region from one to several cd$^{-1}$, many of 
the periodograms steadily rise in a ``1/$f$'' fashion indicative of systematic or ``red'' 
noise trends \citep[e.g.,][]{1992nrfa.book.....P} close to the maximum timescale of one day. 
However, apart from one object (SOri~J053825.4-024241), no potential periodic signals stand out 
high enough above the baseline.  Here we have used a criterion of 4-$\sigma$, equivalent to 
99.9\% certainty, as explained in CH10. Since SOri~J053825.4-024241 
displays intrinsic aperiodic variability at the 10\% level, and the putative 3.5~cd$^{-1}$ 
(or $\sim$9.6h) signal barely exceeds the detection threshold at S/N$\sim$4, it is unclear 
as to whether this is a true periodicity.

\subsection{Prospects for pulsation}

The lack of periodic signals in the 1--4-hour range suggests that none of the $\sigma$~Ori 
cluster members in our sample exhibits deuterium-powered pulsation at a level above several 
millimagnitudes. However, the strength of this conclusion depends on the likelihood that 
one or more targets fall on the PB05's predicted pulsation instability strip. If we assume 
a distance of 350~pc for the $\sigma$~Ori cluster, then all seven BDs in the sample 
(S~Ori~J053825.4-024241, S~Ori~J053826.1-024041, S~Ori~31, S~Ori~J053829.0-024847, 
S~Ori~53, S~Ori~27) may be on the instability strip, to within the uncertainties. If we 
instead adopt a distance of 440~pc, then the VLMS S~Ori~J053826.8-022846 becomes an 
additional candidate, whereas the position of S~Ori~45 falls slightly off the strip. Thus 
one would naively expect that a handful of our targets have temperatures and luminosities 
consistent with those required for pulsational instability. Nevertheless, the significant 
size of the measurement uncertainties compared with the width of the strip must be taken 
into account.

 To estimate the chance that in fact {\em none} of our sample have H-R diagram positions 
overlapping the instability strip, we adopt temperature-luminosity probability 
distributions for each object. We take the distributions to be two-dimensional asymmetric 
gaussians, normalized and centered at the adopted luminosities and temperatures. The 
gaussian widths are given by the associated 1-$\sigma$ uncertainties. The position of each 
target then corresponds to a probability that it is susceptible to pulsation, which we 
determine by integrating its distribution over the entire region of the instability strip. 
For objects on or very close to the strip, this value is $\sim$20-25\%, whereas for the 
higher mass stars far from the strip it is close to zero. The probability that the position 
of a given object does {\em not} overlap with the instability strip then ranges from 75\% 
to 100\%. The product of these values over all targets provides an estimate of the chance 
that no pulsators would be present in our sample. For luminosities derived from $J$-band 
magnitudes and a cluster distance of 440~pc, this probability is 19\%-- small but certainly 
not negligible. Since systematic errors can significantly affect the result, we have also 
recalculated this value using both $I$-band-derived luminosities and the alternate distance 
of 350~pc. The different combinations yield probabilities from 23\% to 32\%. Turning these 
numbers around, there is a $\sim$70-80\% chance that at least one object should exhibit 
pulsation based on its position within the instability strip, assuming that the theoretical 
calculations underpinning it (PB05) do not suffer from gross systematic errors.
 
In addition, the expectation value for the number of objects lying directly on the strip 
lies between one and two, depending on the choice of distance and $J$ or $I$ band 
magnitudes. Therefore if pulsation is operating at an observable level, we are likely to 
detect at least one instance of it. The fact that we also did not detect short-timescale 
variability in our larger ground-based sample (CH10) of BDs and VLMSs suggests that the 
lack of observable pulsation is a repeatable result and not the product of sample selection 
effects. Nevertheless, we caution that the small number of low-mass objects observed here 
leaves open the possibility that no true pulsation candidates were among our sample.

Assuming this is not the case, statistically we expect at least one object in our sample to be 
susceptible to pulsation. If so, the amplitude of this phenomenon appears to be too low to be 
observed in the infrared. To quantify the detection limit, we have fit power laws to each 
periodogram, tracing out the maximum amplitude level as a function of frequency. These curves, 
of form $A/(f+B)+C$ for frequency $f$ and constants $A$, $B$, and $C$, mark the highest 
amplitude signal observed for each object, whether real or noise. We display them in Fig.\ 6, 
where the procedure has been carried out on both the raw and pixel-phased corrected data.

\begin{figure*}
\begin{center}
\epsfxsize=2\columnwidth  
\epsfbox{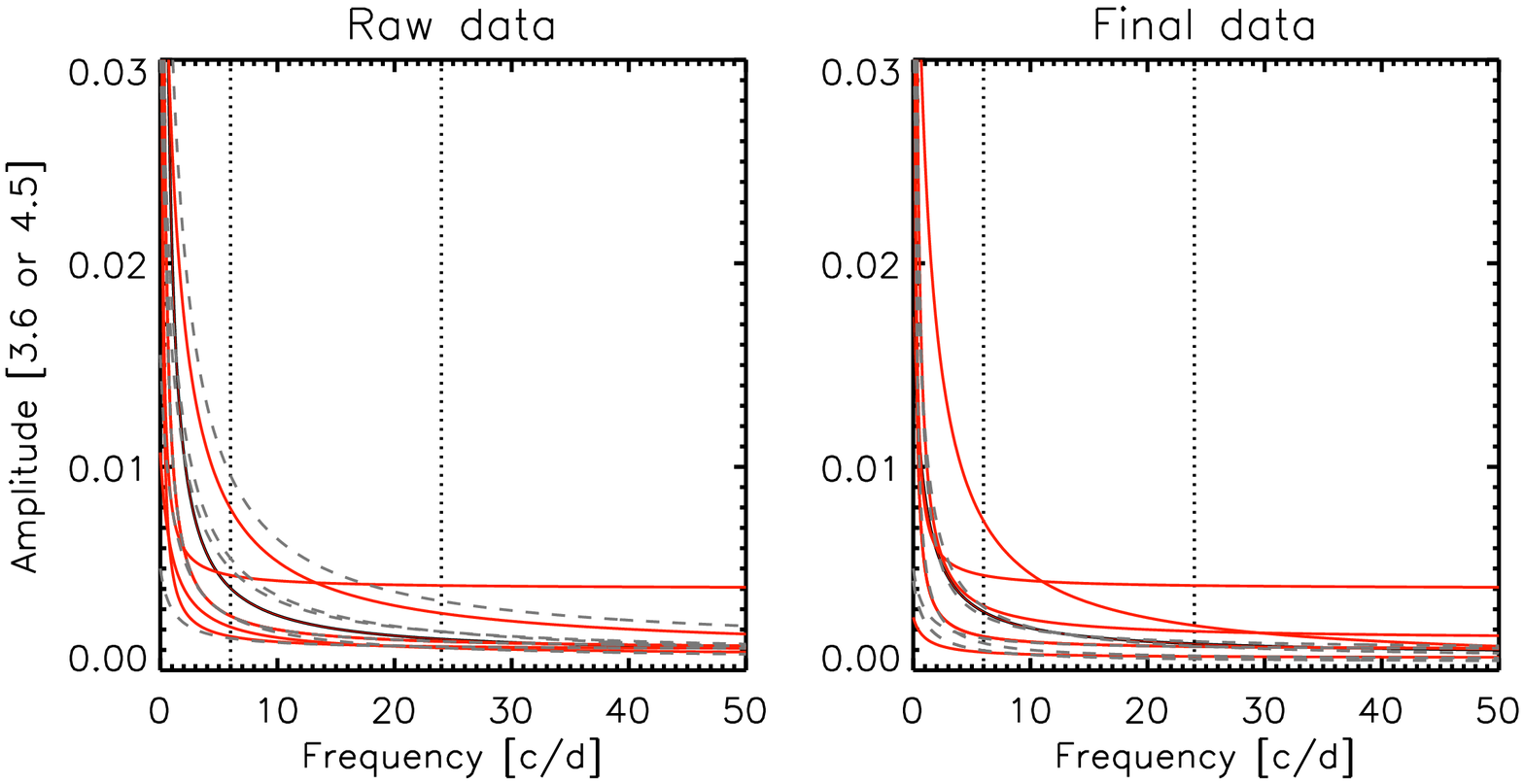}
\end{center}
\caption{Limits on pulsation detection in the periodograms based on raw light curves. Solid red curves
show the limits for objects on or near the instability strip, whereas grey dashed curves are for
objects not expected to exhibit pulsation. The curve for S~Ori~53 is off the chart at a uniform 
amplitude of 0.04 magnitudes. The dotted lines indicate the region of frequency space where we 
expect pulsation to occur (i.e., 1--4-hour periods). The left panel displays data before
pixel-phase correction, and the right shows the final data after removal of pixel-phase 
oscillation trends. For the few objects that did not require these corrections, the curves
derived directly from raw data are repeated.}
\end{figure*}

For the majority of objects, we have detected no periodicities in the pulsation frequency 
range with amplitudes greater than several millimagnitudes. Brown dwarfs S~Ori~45 and 
S~Ori~53 are exceptions, as they have higher limits (0.005 and 0.04 magnitudes respectively 
in the 4.5~$\mu$m band) owing to their faintness and correspondingly high noise levels in 
both the light curves and periodograms. In addition, brown dwarf S~Ori~J053825.4-024241 has 
a higher limit for pulsation (0.004-0.007 magnitudes in the 3.6~$\mu$m band, depending on 
frequency) since it displays substantial intrinsic variability. The rest of our targets 
have maximum amplitudes in the periodogram of at most 0.002 to 0.003 magnitudes. This 
represents the threshold above which we detect no periodicities. We conclude that if 
deuterium-burning pulsation is present in any of our sources (apart from the three 
exceptions noted above), then its amplitude must be below this level.

\subsection{SOriJ053825.4-024241: a high-amplitude variable brown dwarf}

Among our sample, the substellar $\sigma$~Ori member SOriJ053825.4-024241 stands out as the lone 
target highly variable on timescales less than 24 hours. With a 3.6~$\mu$m band RMS of 0.035 
magnitudes, this object has a peak-to-peak amplitude of 0.15 magnitudes. It displays variations 
about four times as large in the $I$-band, based on our longer timescale ground-based dataset 
(CH10). Other studies \citep{2006A&A...445..143C} have indicated that 
SOriJ053825.4-024241 is actively accreting and has a disk \citep{2007ApJ...662.1067H}. 

No previous infrared studies of brown dwarfs have uncovered aperiodic variability on such short 
timescales. However, variability of young stars at Spitzer wavelengths or of brown dwarfs in 
general with these amplitudes and on longer time scales is not unprecedented. The Young Stellar 
Object Variability (YSOVAR) project \citep{2011ApJ...733...50M} campaign on young Orion Nebula 
Cluster stars (masses $\gtrsim$0.1~$M_\odot$) has also found substantial erratic variability in 
the 3.6 and 4.5~$\mu$m bands. Assessment of their data has shown that the aperiodic variables 
among the sample known to harbor disks display a range of variability RMS values centered on 
$\sim$0.03 magnitudes in the 3.6~$\mu$m band (Morales-Calder\'{o}n 2011, private communication). 
Similar amplitude distributions were obtained using existing 
multi-epoch data with limited cadence in Taurus and Chamaeleon~I by \citet{2008ApJ...675.1375L} 
and \citet{2010ApJS..186..111L}. The typical RMS of a few hundredths of a magnitude is quite 
consistent with the value that we have measured for S~Ori~J053825.4-024241. 
\citet{2011ApJ...733...50M} discuss the possible causes of the mid-infrared variability and 
surmise that many of their variables may be explained by variable obscuration by overdense 
regions in the inner disk, while others are caused by intrinsic changes in the inner disk 
emission itself. Either of these scenarios may apply to SOriJ053825.4-024241. In any case, hot 
accretion gas is likely too faint at infrared wavelengths to serve as the source of variability 
for this object.

To further explore the behavior of this BD on different timescales, we have performed an 
autocorrelation analysis. In addition to displaying quasi-periodicity patterns not picked up by 
the periodogram, it is useful in assessing the timescale on which the variability mechanism 
remains coherent. We have calculated an autocorrelation function based on the 
S~Ori~J053825.4-024241 light curve using both a standard, ``biased'' formula, as well as one that 
corrects for the finite data length. The standard autocorrelation function (ACF) is given by: 
$$A(t)=\frac{1}{A(0)}\sum_{j=1}^{N-t/\Delta t}L(j)L(j+t/\Delta t),$$ where $L(j)$ are the 
light curve points, $\Delta t$ is the time spacing between datapoints (which must be uniform), $N$ 
is the total number of points, and the $A(0)$ factor in front in included so that at a time lag 
of zero, the ACF is completely correlated ($A(0)=1$).

To account for the fact that fewer points are available to calculate the ACF at longer lag times 
($t>0.5$), we have produced another version- the ``unbiased'' ACF- in which the this roughly linear 
effect ($\sim N-t/\Delta t$)) has been divided out. In both cases, we have computed the autocorrelation 
via fourier transform of the power spectrum \citep[as specified by the Wiener-Khinchin 
theorem;][]{wiener,khinchin}, since this is both faster and less prone to numerical inaccuracies.

Both versions of the ACF are plotted in Fig.\ 7. We find that the light curve is well correlated 
up to timescales of $\sim$0.15d, or 3.6h. At longer timescales, it also shows significant 
correlation due to the overall trend seen in the light curve; this is illustrated by the two 
peaks at $\sim$0.43d and $\sim$0.9d (the latter primarily in the unbiased ACF). We conclude that 
the variability mechanism is physically coherent on timescales of at least a few hours. The 
hypothesis of variable obscuration in association with the disk is qualitatively consistent if 
the scale of clumpiness and location of dust is such that fluctuations would pass by the face of 
the BD in several hours.

\begin{figure}
\begin{center}
\epsfxsize=.99\columnwidth 
\epsfbox{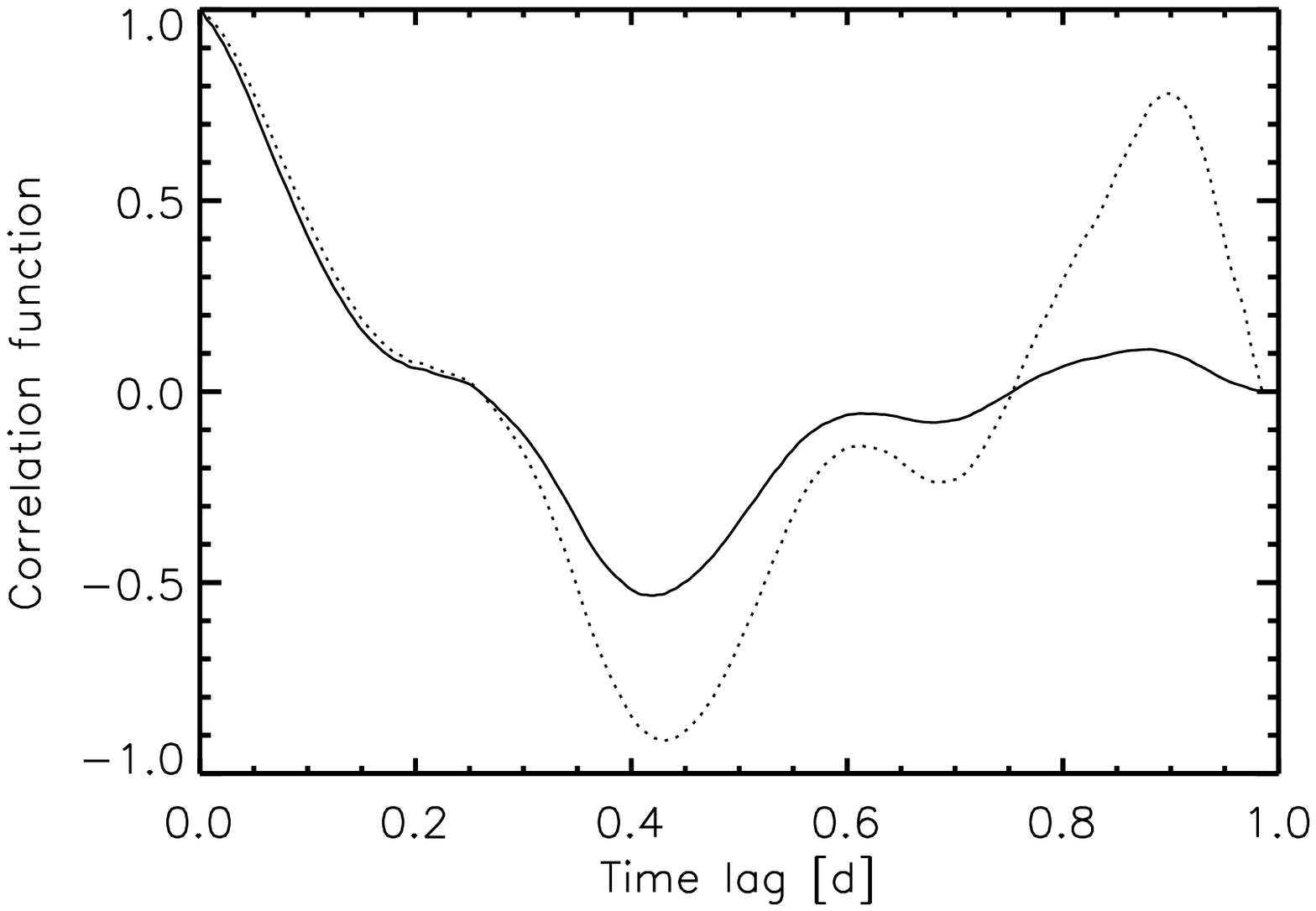}
\end{center}
\caption{Autocorrelation functions for SOriJ053825.4-024241. The solid line is the standard ACF, 
whereas the dashed line is the version that has been corrected for finite data length.}
\end{figure}

\section{Comparison with optical data}

Previously we observed a region including both Spitzer fields, using the CTIO 1.0-meter telescope 
(CH10). High cadence (every 10 minutes) data in the Cousins $I$ band and lower cadence (twice per 
night) data in $R$ was acquired over runs of 12 and 11 consecutive nights, respectively in 2007 and 
2008. We identified a number of variable objects, both in the $\sigma$~Ori cluster and background 
field. Although the overall time baseline of the {\em Spitzer} observations is short compared to that 
of the ground-based campaign, we have searched for common variability in the two datasets.

A total of seven variable $\sigma$~Ori cluster members from the ground-based campaign fall in the 
fields of the {\em Spitzer} observations, as noted in Table~1. In the 4.5~$\mu$m field, 
S~Ori~J053833.9-024508 and 4771-41 were identified as aperiodic variables in the ground-based 
photometry. In addition, the BD S~Ori~45 was identified as being periodic in the $I$-band, with a 
period of 7.2 hours and amplitude 0.034 magnitudes, whereas VLMS SWW40 was found to have a period 
of 4.47 days and amplitude 0.013 magnitudes. In the 3.6~$\mu$m field, the BD S~Ori~J053825.4-024241 
was identified as an aperiodic variable, as discussed in $\S$5.2. Two additional variables were 
found be periodic in the $I$ band: S~Ori~J053823.6 ($P$=1.7d; $A$=0.017 mags) and 
S~Ori~J053817.8-024050 ($P$=2.4d; $A$=0.008 mags).

For those ground-based variables with brightness fluctuations on timescales longer than a day, we 
do not necessarily expect to observe variability in our shorter {\em Spitzer} dataset. Indeed, we 
do not recover periodic variability at great than the 1\% level in any of the ground-based 
periodic variables. In addition to the shorter time baseline, it is possible that the 
non-simultaneity of observations and the different wavelengths make rotational spot modulation-- 
the primary explanation for periodic variability in young VLMSs and BDs-- unobservable in our 
light curves.

Several of the previously identified aperiodic variables, on the other hand, do appear to be 
variable at infrared wavelengths. The BD S~Ori~J053825.4-024241 displays relatively high 
amplitude erratic fluctuations (see $\S$5.2). Object 4771-41 shows residual variability after 
correction for the pixel-phase effect (light curve RMS of $\sim$0.01 versus $\sim$0.001 
magnitudes), and S~Ori~J053833.9-024508 also displays variability at a significantly higher 
level than predicted by signal-to-noise estimates (light curve RMS of $\sim$0.05 versus $<$0.01 
magnitudes). The RMS values in the {\em Spitzer} bands are similar to those found in the 
optical for S~Ori~J053825.4-024241 and S~Ori~J053833.9-024508, whereas they are roughly an 
order of magnitude lower for 4771-41. Thus the light curve of this latter object may exhibit 
residual pixel-phase effects, as opposed to real variability. However, for the other two 
aperiodic variables, the rough correspondence of RMS amplitudes in both the optical and 
infrared suggests that the variability mechanism may be relatively insensitive to wavelength.

Interestingly, object S~Ori~J053829.0-024847 displays substantial variability at 4.5~$\mu$m (an 
0.06-magnitude drift over 24 hours), whereas it did not appear variable in our ground-based 
dataset. We suspect that the variability mechanism in this case was dormant during the optical 
observations, although its photometry could have been affected by a nearby neighbor on the array. 
Since this object exhibits an infrared excess \citep{2007ApJ...662.1067H,2007A&A...470..903C}, 
there is an additional possibility that the variability is associated with the disk and thus only 
visible in the near-infrared and at longer wavelengths.

In addition to the recovery of aperiodic variability in our $\sigma$~Ori cluster sample, we 
also re-identify a number of eclipsing binaries; further details on these field objects are 
provided in the appendix.

\section{Conclusions}

We have presented high-cadence infrared light curves of 14 low-mass $\sigma$~Orionis cluster members 
based on Warm {\em Spitzer} observations. The excellent precision of our photometry led to limits of 
0.002--0.003 magnitudes on the amplitude of any deuterium-burning pulsation in these objects in the 
3.6 and 4.5~$\mu$m bands. This result is consistent with our ground-based $I$-band findings, which 
revealed no periodicities with amplitudes greater than 0.01 magnitudes and timescales shorter than 7 
hours among low-mass $\sigma$~Ori cluster members. In this work we have reduced the amplitude limit 
by an order of magnitude, albeit on a smaller sample of BDs and VLMSs. Notably, we also find little 
other variability on the $<$24-hour timescales probed by the data. The main exception was brown 
dwarf 
S~Ori~J053825.4-024241, which displays brightness variations of up to 0.1 magnitudes over the course 
of a day. The similarity of amplitudes in the infrared and optical suggests that obscuration by dust 
material in the surrounding disk provides a better explanation than does variable light scattering or 
accretion \citep[e.g.,][]{2010A&A...517A..16V}. We propose that a general lack of variability 
among young, low-mass cluster members may in fact be useful for future studies at infrared 
wavelengths, such as searches for planets around young BDs and VLMSs.  Both transit detection and 
radial velocity measurements benefit from low levels of spot-related activity on short timescales.

We also emphasize that production of light curves devoid of the pixel-phase and other detector 
effects is difficult at present with Warm {\em Spitzer} data. While previous work has 
successfully identified low-amplitude planetary transit signatures in {\em Spitzer} 
light curves \citep[e.g.,][]{2011ApJ...726...95D}, the transit event represents only a small portion 
of these time series and hence systematic trends can be taken into account. When
the {\em entire} light curve is instead the subject of interest, and the form of
variability is unknown in advance, systematics are more difficult to model and remove. For future 
high-precision photometric time series work, we recommend further exploration of the sensitivity 
distribution within individual pixels, perhaps through even higher cadences that might provide 
more datapoints over a given time and thus greater spatial coverage within individual pixels.

\acknowledgements{This work is based on observations made with the Spitzer Space Telescope, which is 
operated by the Jet Propulsion Laboratory, California Institute of Technology under a contract with 
NASA. We acknowledge support from NASA under contract 1382589 administered through JPL/Caltech.
We thank the referee, Kevin Luhman, for helpful comments. We also express appreciation to Sean Carey and 
Roberta Paladini for help in correcting pull-down and bias residuals, as well as Mar\'{i}a 
Morales-Calder\'{o}n for extensive advice on Warm {\em Spitzer} reductions.}

\clearpage

\section*{Appendix}

In addition to examining the light curves of the 14 $\sigma$~Ori cluster member targets, 
we also searched the entire 3.6 and 4.5~$\mu$m fields for serendipitous 
foreground and background variables. After producing light curves for all point sources 
with magnitudes less than $\sim 19.0$, we assessed their RMS spread as a function of brightness. 
Objects lying more than three standard deviations above the median trend were flagged as 
possible variables. We visually examined their light curves and disregarded those whose 
brightness fluctuations were clearly caused by pixel sensitivity effects. Four objects 
(other than BD 053825.4-024241; $\S$5.2) displayed conspicuous variability by these criteria; 
their light curves are presented in Fig.\ 8. For consistency with the other presented light 
curves, we show both the time series and their periodograms. We list the estimated period, which 
often does not correspond to the largest periodogram peak since this analysis method is 
relatively insensitive to the presence of secondary eclipses.

All four stars were also identified as variables in our $I$-band ground-based dataset 
(CH10); therefore, we refer to them by the same nomenclature. We have not rigorously fit eclipse 
profiles or other models to the data but present estimates ($\sim$10--20\% accuracy) of their 
main parameters here:

{\bf CTIO~J05381870-0246582} is an eclipsing binary system with an
$I$-band depth of $\sim$0.45 magnitudes, and 4.5~$\mu$m depth of at least 1.2 magnitudes. The 
most likely period is $\sim$11.8 hours, or 5.9 hours if all of the eclipses are primary (the data 
are too noisy to to distinguish different depths in subsequent eclipses).

{\bf CTIO~J05382129-0240318} also appears to be an eclipsing binary, with period $\sim$9.6 
hours. This period is fully consistent with our ground-based data, for which we 
unfortunately reported an erroneous value (4.6 days instead of 9.5 hours). The 3.6~$\mu$m depth 
($\gtrsim 1.3$ magnitudes) is significantly deeper than the $I$-band depth ($\sim$0.35 
magnitudes). 

{\bf 2MASS~J05382188-0241039} exhibits a slightly asymmetric periodic profile, reminiscent of an 
RR~Lyrae star. Its period of 11.8h is also consistent with this type of pulsator. Since the 
timescale is so close to half a day, aliasing caused us to misidentify and report a 1.0d period 
for the ground-based data. The 3.6~$\mu$m peak-to-peak amplitude is $\sim$0.25 magnitudes, whereas 
the value at $I$ band is just over 0.6 magnitudes.

{\bf 2MASS~J05381949-0241224} also displays the characteristic shape of a close eclipsing binary, 
although there is slight decrease in its peak amplitude over 24 hours which may be attributed to 
systematic pixel sensitivity effects. The period is 2.8 or 5.6 hours, depending on whether 
alternating brightness dips are secondary eclipses. At $\sim$0.5 magnitudes, the peak-to-peak 
amplitude at 3.6~$\mu$m is about 20\% smaller than that in the $I$ band.

\begin{figure*}
\begin{center}
\plottwo{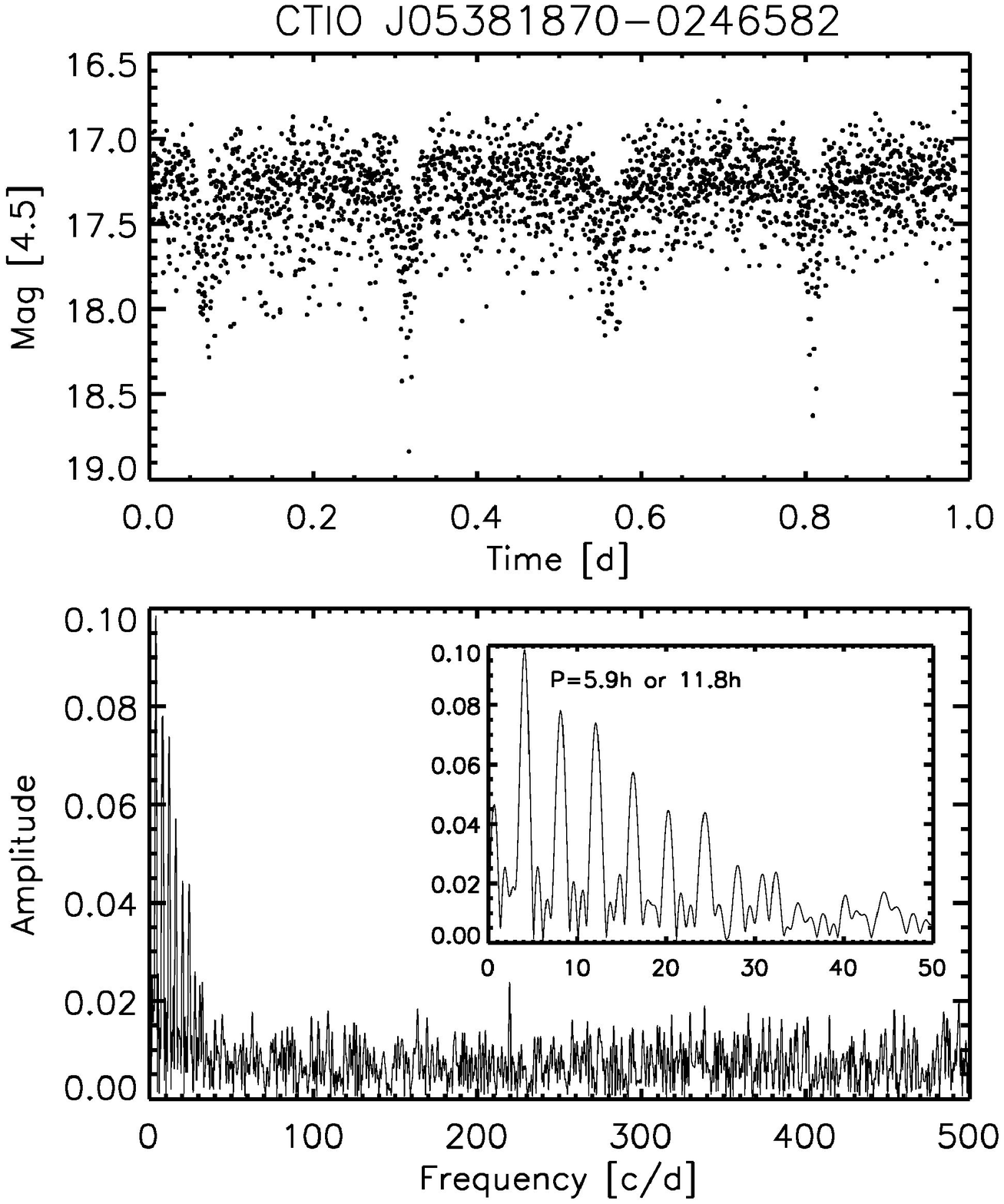}{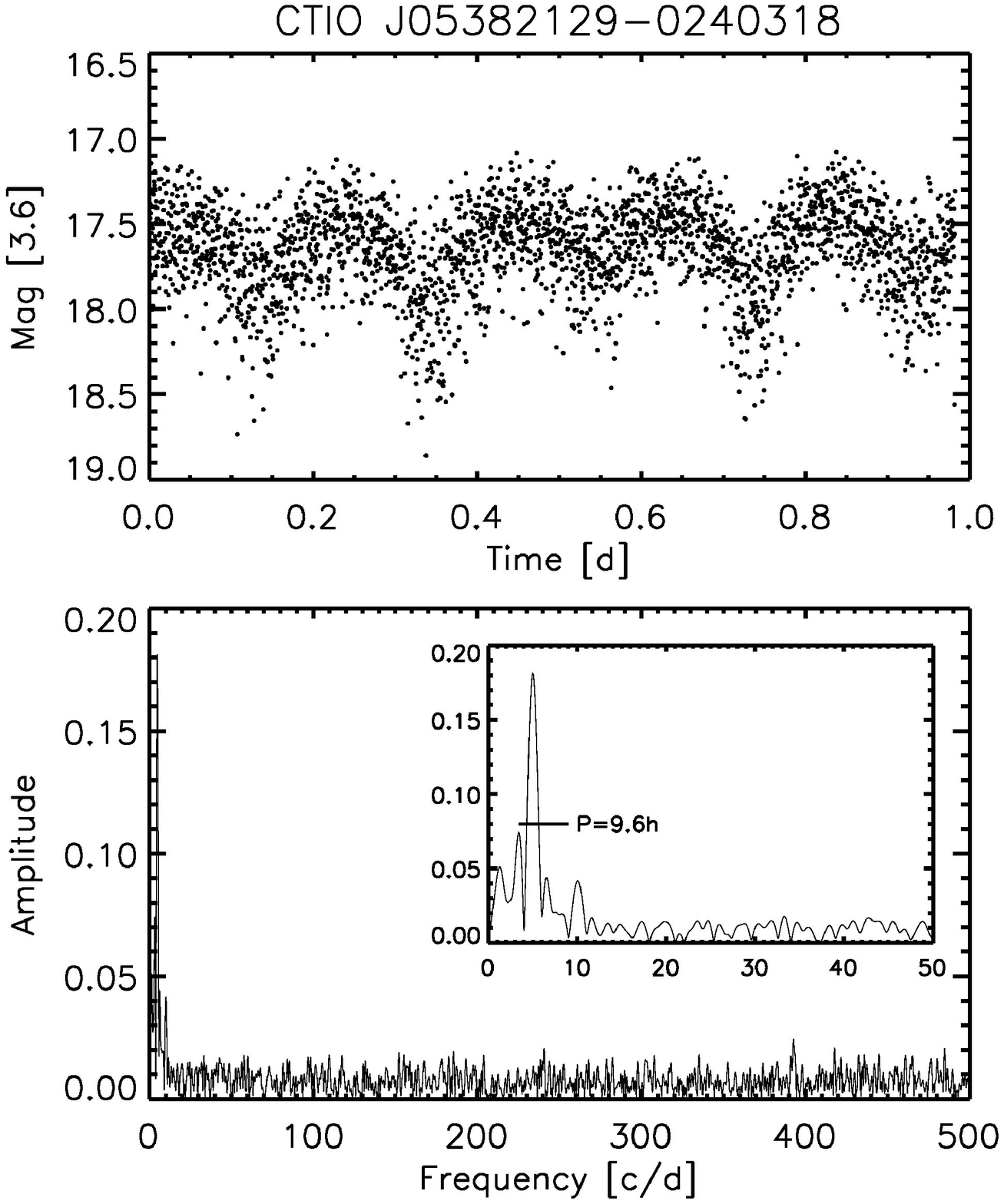}
\end{center}  
\caption{Field variable stars. Light curves (top) and periodograms (bottom) are as in Fig.\ 5; 
estimated periods are marked near the corresponding frequency peaks.}
\end{figure*}  

\addtocounter{figure}{-1}
\begin{figure*}
\begin{center}
\plottwo{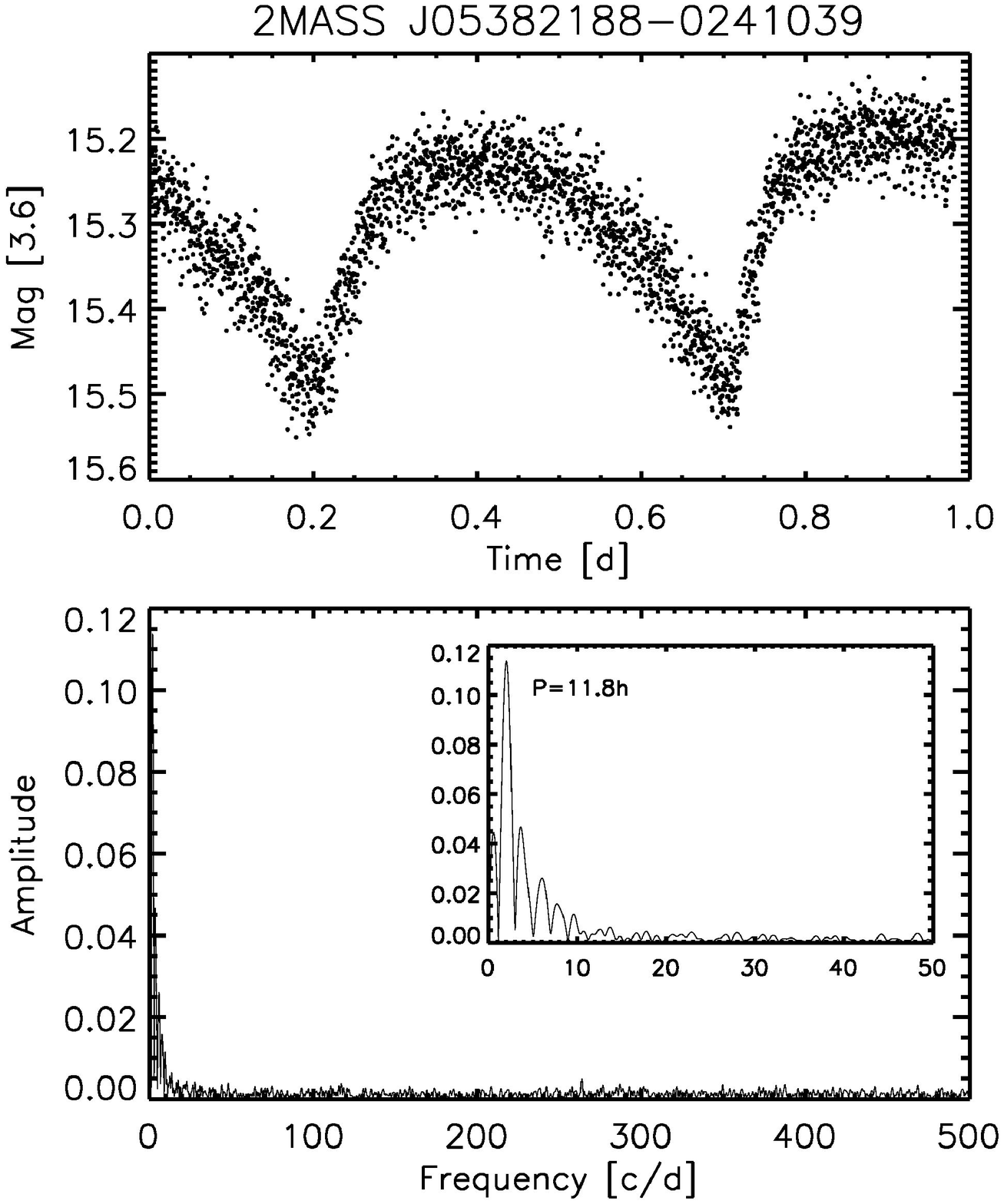}{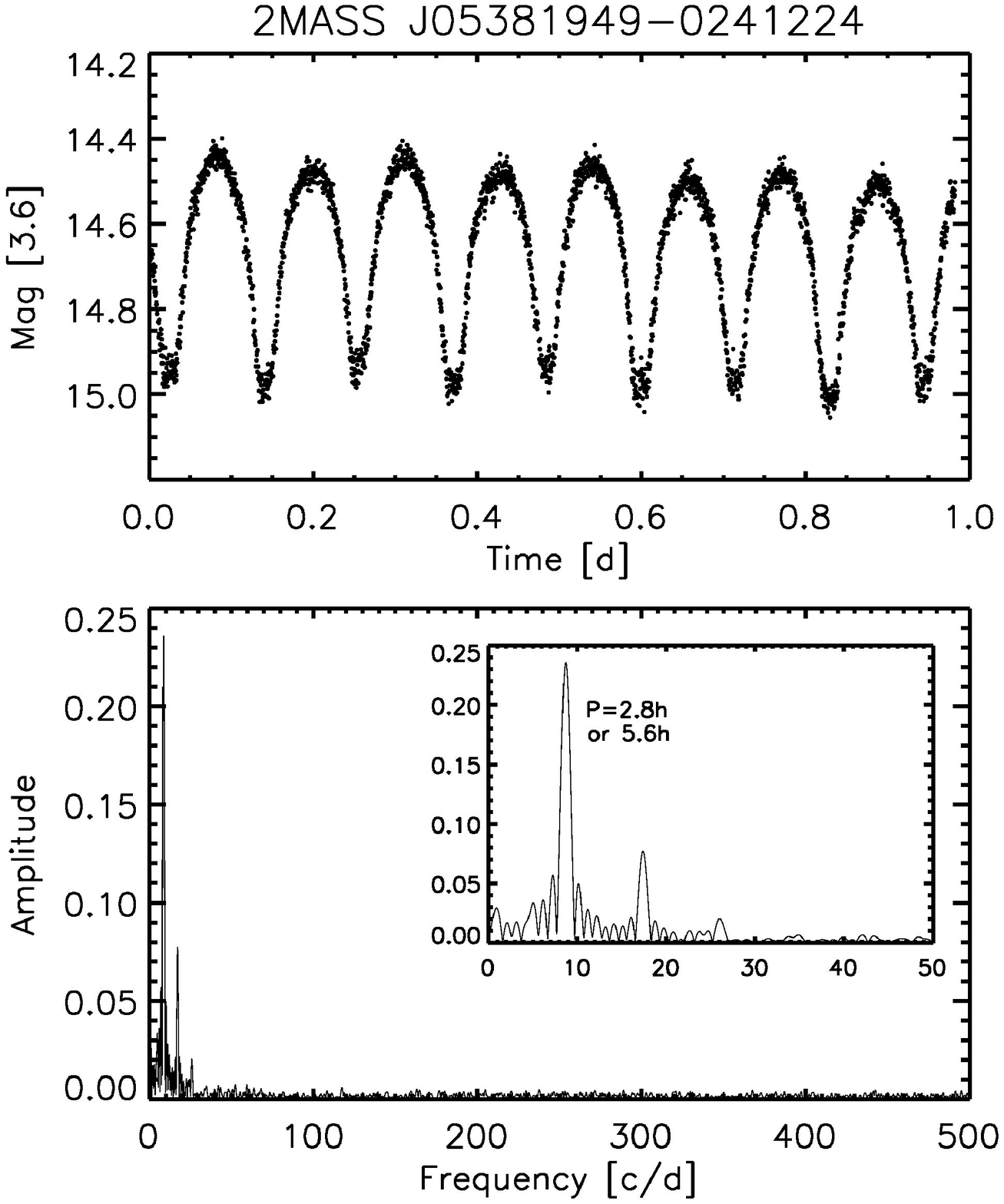}
\end{center}
\caption{(cont.)}
\end{figure*}

\newpage
\bibliographystyle{apj}
\bibliography{Spitzer}

\end{document}